\documentclass[twocolumn,aps,prxquantum,reprint,showpacs,superscriptaddress,longbibliography]{revtex4-1}

\usepackage{amsmath,amsfonts,amsthm,amssymb}
\usepackage{graphicx}
\usepackage{dcolumn}
\usepackage{bm}
\usepackage[unicode=true, breaklinks=false, pdfborder={0 0 1}, backref=false, colorlinks=true, linkcolor=blue, citecolor=blue, urlcolor=blue]{hyperref}
\usepackage{color}
\usepackage{xcolor}
\usepackage{mathrsfs}
\usepackage{float}
\usepackage{txfonts}
\usepackage{wasysym}
\usepackage{enumitem}

\newcommand{\marti}[1]{{\color{blue} #1}}
\newcommand{\pablo}[1]{{\color{purple} #1}}

\makeatletter

\newcommand{\Rmnum}[1]{\expandafter\@slowromancap\romannumeral #1@}
\makeatother
\begin{document}
\title{
Geometric optimization of non-equilibrium adiabatic thermal machines
\texorpdfstring{\\}{ }
and implementation in a qubit system}

\author{Pablo Terr\'en Alonso}\thanks{These two authors contributed equally.}
\affiliation{International Center for Advanced Studies, Escuela de Ciencia y Tecnolog\'ia and ICIFI, Universidad Nacional de San Mart\'in, 
Avenida 25 de Mayo y Francia, 1650 Buenos Aires, Argentina\:}

\author{Paolo Abiuso}\thanks{These two authors contributed equally.}
\affiliation{ICFO—Institut de Ci\'encies Fot\'oniques, The Barcelona Institute of Science and Technology, 08860 Castelldefels (Barcelona), Spain\:}
\affiliation{D\'epartement de Physique Appliqu\'ee, Universit\'e de Gen\'eve, 1211 Gen\'eve, Switzerland\:}

\author{Mart\'i  Perarnau-Llobet}
\affiliation{D\'epartement de Physique Appliqu\'ee, Universit\'e de Gen\'eve, 1211 Gen\'eve, Switzerland\:}

\author{Liliana Arrachea}
\affiliation{International Center for Advanced Studies, Escuela de Ciencia y Tecnolog\'ia and ICIFI, Universidad Nacional de San Mart\'in,  
Avenida 25 de Mayo y Francia, 1650 Buenos Aires, Argentina\:}
\date{\today}

\begin{abstract}
We adopt a geometric approach to describe the performance of adiabatic quantum machines, operating under slow time-dependent driving and in contact to two 
reservoirs with a temperature bias during all the cycle. We show that the problem of optimizing the power generation of a heat engine and the efficiency of both the heat engine and refrigerator operational modes
is reduced to an isoperimetric problem with non-trivial underlying metrics and curvature. This corresponds to the maximization of the ratio between the area enclosed by a closed curve and its corresponding  length. We illustrate this procedure in a qubit coupled to two reservoirs operating as a thermal machine by means of an adiabatic protocol. 
\end{abstract}
\maketitle
\section{Introduction}

The development and implementation of thermodynamic processes in few-level quantum systems is currently a very active area of research. Thermodynamic cycles  conceived for  macroscopic working substances (WS), such as the Otto or Carnot cycle, are now realized 
in  single atoms~\cite{pekola2019mar,von2019spin,rossnagel2016apr,ronzani2018oct,deassis2019jun,peterson2019dec}
%
%
and large theoretical efforts are devoted to its characterization and optimization at the microscopic scale \cite{rezakhani2009aug,Schmiedl2007,sivak2012may,Kosloff2017,paolo,paolo2,abiuso2019non,cavina2021maximum,adiageo}.
In these  standard thermodynamic cycles, the WS operates in four steps, of which two are in contact to reservoirs at different  temperatures connected one at a time, while  the other  two steps consist in an evolution decoupled from the reservoirs.  
It is however typically hard to fully isolate a quantum WS from the environment, which is required to emulate  ideal classical cycles. This motivates the study of non-equilibrium systems, where the driven WS is permanently in contact with two or more reservoirs.  Unlike  standard  thermodynamic cycles, these microscopic machines operate away from equilibrium during all the cycle. Thermoelectric devices \cite{Benenti2017} as well as autonomous refrigerators \cite{Palao2001,Youssef2009,Brunner2012} are seminal examples of this type of operation. 

When the WS is connected 
at the same time to two or more thermal reservoirs, it is permanently thread by a heat flux. Hence, the very operation as a machine relies on the mechanism of heat--work conversion
in order to overcome this effect as well as the dissipation generated by the driving sources. The optimal machine  is the one leading to the optimal balance between these two processes.
In quantum systems, the operation under a small temperature bias and
``adiabatic driving" through parameters which slowly vary on time is of paramount relevance, since this is an appealing scenario to control the non-equilibrium mechanisms. 
In this regime, the period of the cycle is larger than any characteristic time of the quantum system, including the relaxation time between system and reservoirs~\cite{thouless1983may,brouwer1998oct,zhou1999jan,moskalets2002nov,moskalets2004dec,reckermann2010jun,cavina2017slow}. 
%
%

Recently, it was proposed that the dissipation and the heat--work conversion mechanisms are respectively described by different components of the thermal geometric tensor. Furthermore,
  the heat--work conversion component can be expressed in terms of a Berry-type phase~\cite{adiageo}, which has an associated Berry-type curvature~\cite{berry1984quantal}, and similar ideas were followed in \cite{Hino2021,Izumida2021}.
Hence,  a length and an area in the parameter space can be defined. Besides, it is well known that dissipation and entropy production admit a geometric description in terms of the concept of thermodynamic length~\cite{weinhold1975sep,salamon1980jul,salamon1983sep,nulton1985jul,schlogl1985dec,andresen1996nov,diosi1996dec,
crooks2007sep,campisi2012,VanVu2021}. This geometric approach has proven useful to optimize finite-time thermodynamic processes (examples can be found in ~\cite{sivak2012may,Zulkowski2012,Zulkowski2013,bonanca2014jun} for classical and~\cite{zulkowski2015sep,scandi2019oct,paolo2} for quantum systems), including  the finite-time Carnot cycle \cite{paolo,paolo2} and slowly driven engines~\cite{brandner2020jan,millermoha,millerTUR,frim2021optimal,topical}.
As mentioned before, these cycles are characterized by the WS being coupled to a single reservoir or completely decoupled from reservoirs.

The aim of the present work is to optimise the performance of thermal machines with cycles in permanent contact to two or more reservoirs at different temperatures by a geometrical approach. To this end we combine the geometrical description of the two competing mechanisms of the non-equilibrium thermal machine (namely heat-work conversion and dissipation) in order to find optimal protocols for maximizing power generation of the heat-engine operation and the efficiency of the heat engine and refrigerator operational modes. We show that the problem of finding such optimal protocols  reduces to an \emph{isoperimetric problem}~\cite{ros2001isoperimetric} (also studied as \emph{Cheeger Problem}~\cite{parini2011introduction,leonardi2015overview}), that is the task of finding the shape which maximizes the ratio between area and length. This is one of the oldest geometric problems in history, and was solved already by the ancient Greeks in the standard 2-dimensional Euclidean plane~\cite{blaasjo2005isoperimetric}. Nevertheless, when the underlying area density or length metrics are nontrivial~\cite{howards1999isoperimetric,morgan2005manifolds,rosales2008isoperimetric,carroll2008isoperimetric}, no general solution is known.

We illustrate these ideas in a prominent quantum system playing the role of the WS: a qubit driven by two parameters slowly changing in time and asymmetrically coupled to two thermal reservoirs at different temperature (see Fig.~\ref{fig:pic_model}).  We show analytically that the limiting value for the area in the parameter space is given by the celebrated
Landauer  bound~\cite{landauer61,landauer88}, which has been the motivation of many studies including several experiments (see e.g. \cite{berut2012experimental,jun2014high}).  
We also find that, operating as a heat engine, the qubit thermal machine offers a very good ratio between generated power and efficiency in a wide range of parameters.

The paper is structured as follows. In Sec.~\ref{sec:setup}, we introduce the set-up and define the relevant thermodynamic quantities  to characterize the cycle. In Sec.~\ref{sec:geom-opt}, we describe the underlying geometry of the system. In Sec.~\ref{sec:timeopt} we describe the heat engine and refrigeration modes of the machine, and perform the optimization with respect to the driving time. In Sec.~\ref{sec:results}, we develop the full optimization of the machine. We then compute in detail all the relevant quantities in a model of one of the most paradigmatic and simplest quantum engines, namely a driven qubit system (see Refs.~\cite{karimi2016nov,paolo2,adiageo}). 

\section{The setup and its thermodynamics}
\label{sec:setup}
\begin{figure}
    \centering
    \includegraphics[width=0.5\textwidth]{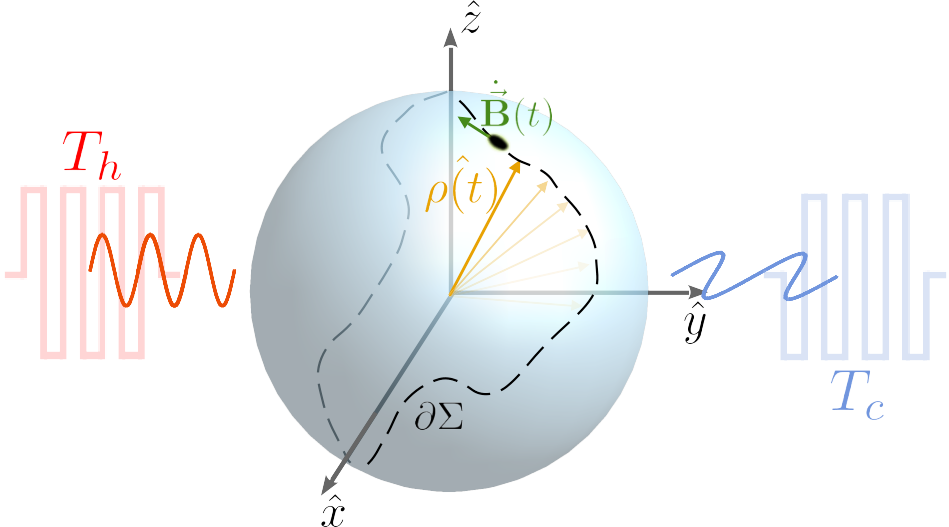}
    \caption{Schematic configuration of the setup. A working substance WS is in contact with two reservoirs at different temperatures, $T_c$ and $T_h$. The state $\hat \rho$ of the system changes at slow but finite speed along a closed path defined by the Hamiltonian $\mathcal{H}(\vec B(t))$ in a quasistatic process.
    }
    \label{fig:pic_model}
\end{figure}

We focus on the usual configuration where the WS operates in contact to two reservoirs at different temperatures $T_{\rm h}$ (hot) and $T_{\rm c}$ (cold), with  $T_{\rm h}=T+\Delta T$ and $T_{\rm c} \equiv T$. A particular example, which will be studied in detail in forthcoming sections is sketched in Fig.~\ref{fig:pic_model}.
The full system is described by the Hamiltonian 
 \begin{equation}
  {\cal H}(t)= \sum_{\alpha={\rm c,h}} \left( {\cal H}_{\alpha} + {\cal H}_{\rm cont,\alpha} \right) + {\cal H}_{\rm WS}(t).
  \label{hamtot}
  \end{equation}
The Hamiltonian for the WS depends on time through a set of control parameters $B_j(t),\; j=1,N$, which we enclose in a vector $\vec{ B}(t)=\left(B_1(t), \ldots, B_N(t)\right)$. Hence,
${\cal H}_{\rm WS}(t)={\cal H}_{\rm WS}(\vec{ B}(t))$. We are interested in cycles, so that we consider time-dependent protocols satisfying $\vec{ B}(t+\tau)= \vec{B}(t)$, being $\tau$ the period of the cycle.
The reservoirs are represented by the Hamiltonian
\begin{equation}\label{qres}
{\cal H}_{\alpha}=\sum_{k}\varepsilon_{k\alpha}b_{k\alpha}^\dagger b_{k\alpha}, \;\;\; \alpha={\rm c, h},
\end{equation}
 with $b_{k\alpha}$ and $b_{k\alpha}^\dagger$ being the annihilation and creation operators of a bosonic excitation.
The coupling is represented by
\begin{equation}\label{qcont}
{\cal H}_{\rm cont,\alpha}=\sum_{k}V_{k\alpha}\hat{\pi}_{\alpha}\Big(b_{k\alpha}+b_{k\alpha}^\dagger\Big),
\end{equation}
where $\hat{\pi}_{\alpha}$ is a matrix with the dimension of the Hilbert space of the WS.

The crucial concepts that characterize the operation of the thermal machine are the work performed and the net heat  exchanged between the two reservoirs during the cycle. The operation of the driven quantum system as a thermal machine in the presence of a temperature bias  relies on the mechanism of heat--work conversion. 
In the present case we make two main assumptions: \begin{enumerate}[label=(\roman*)]
    \item slow driving \cite{cavina2017slow}, characterized by a small rate of change of the driving parameters with time, $d_t \vec{B}$ (short for $\frac{d}{dt}\vec{B}$), as well as
    \item a small temperature bias $\Delta T$ between the two reservoirs.
\end{enumerate}  
This enables us to work in the linear-response regime with respect to $d_t \vec{B}$ and $\Delta T$. 

A natural theoretical framework in this context is the adiabatic linear response theory 
proposed in Ref.~\cite{ludovico2016feb} in the geometric perspective of Ref.~\cite{adiageo}. This formalism applies to the regime where the period of the cycle is much larger than the longest time-scale characterizing the WS coupled to the reservoirs. In most of the cases, such time scale is determined by the relaxation time $\tau_{\rm rel}$ of the WS with the reservoirs.
More precisely, the dynamical perturbation to the steady state $\hat{\rho}_B$ (corresponding to no driving, i.e. ``frozen" value of $B$), can be estimated to $\delta\hat{\rho}\sim\tau_{\rm rel}(\partial_B \hat{\rho}_{B}) d_t B$ (cf.~\cite{cavina2017slow,ludovico2016feb,adiageo} or Appendix~\ref{appA:reducedDensityMatrix}). Hence, this approach is useful when $\tau\gg \tau_{\rm rel}$. We also consider small temperature bias, such that $\Delta T/T\ll 1$. This description leads to a linear relation between the relevant energy fluxes operating the cycle and the components of the vector 
$d_t{\bf X}=(d_t \vec{B},\Delta T/T)$. The relevant quantities are the net {\em output work} 
and {\em transferred heat} between the hot and cold reservoirs. They are, respectively, defined as the average over one period  of the power developed by the driving sources, and the energy flux into the $\alpha$-reservoir,
\begin{eqnarray}
W &=&
-\int_0^{\tau}dt\;\langle \frac{\partial {\cal H}_{\rm WS}}{\partial \vec{B}}\rangle \cdot d_t {\vec{B}}, \\
Q_\alpha &=& 
-\frac{i}{\hbar}\int_0^{\tau} dt\; \langle \left[{\cal H}_{\alpha},{\cal H}\right] \rangle\;, 
\end{eqnarray}
where $\langle O \rangle=\mbox{Tr}\left[\rho O\right]$, being $\rho$ the global state of system and baths (which in general will be correlated due to the contacts). The corresponding expectations values are evaluated in linear response with respect to $d_t{\bf X}$. In such regime $Q_{\rm c}=- Q_{\rm h}\equiv Q$~\footnote{While the heat flux at each reservoir contains
both transported and dissipated components, the latter contributes at the second order in $d_t{\bf X}$~\cite{adiageo}, as explicitly shown in Eq. (\ref{w}).}.
The result is 
\begin{align}
    W &=\frac{\Delta T}{T}\int_0^\tau dt\;\Vec{\Lambda}\cdot d_t \vec{B} -\int_0^\tau dt\; d_t\vec{B} \cdot \underline{\Lambda}\cdot d_t \vec{B} \;,
   \label{w}\\
    Q &=\int_0^\tau dt\;\Vec{\Lambda}\cdot d_t \vec{B}+\frac{\Delta T}{T}\int_0^\tau\ dt\; \kappa \;.
   \label{q}
\end{align}
These expressions can be derived in the adiabatic linear-response regime as from Ref.~\cite{adiageo} and we defer the reader to that paper for further details.
For the moment it is enough to stress that \{$\underline{\Lambda}$, $\vec{\Lambda}$, $\kappa$\} are all local functions of $\vec{B}$, while they also depend on the coupling parameters, the density of states of the thermal baths and $T$.

In Eq.~(\ref{w}), the first term represents the mechanism of {\em heat--work} conversion and the second one corresponds to finite-time dissipation developed by the time-dependent controls. 


Moreover, in Eq.~\eqref{q}, the transferred heat $Q$  also contains two terms associated to two different physical processes. The first one describes the heat exchange between the reservoirs related to the driving while the second one is the heat transport as a response to the temperature bias. 
Notice that the fundamental component for the thermal machine to operate is the heat--work conversion term $\int_0^\tau dt\;\Vec{\Lambda}\cdot d_t \vec{B}$. In fact, without this component, the only surviving processes are the dissipation of the energy supplied by the driving forces and the trivial conduction of heat as a response to the thermal bias.

The different terms in Eqs.~(\ref{w}) and (\ref{q}) can be reinterpreted geometrically, as explained in the following Sec.~\ref{sec:geom-opt}. This allows for the optimization of the thermodynamic protocols in terms of clear geometrical quantities.

It is important to notice that the second terms of Eqs.~(\ref{w}) and (\ref{q}) have a defined sign. 
In our convention, $\underline{\Lambda}$ is positive definite since it is directly related to the entropy production rate~\cite{adiageo},   which means that it is detrimental for the work output. Similarly, $\kappa$ can be seen to be positive, as a consequence of the fact that this component of the transferred heat describes the flux from the hottest to the coldest reservoir. These are direct consequences of the second law of thermodynamics. Instead, the line integral $\int_0^\tau dt\;\Vec{\Lambda}\cdot d_t \vec{B}$ may have any sign, depending on the driving protocol and it is enough to time-reverse the function $\vec{B}(t)$ to flip the sign. As mentioned before, this term describes the {\em heat--work} conversion process and its sign defines the type of operation of the machine.
In fact, when it is negative, the contribution of the first term of Eq.~(\ref{q}) may overcome the heat flowing into the coldest reservoir and enable the operation of the machine as a \emph{refrigerator}. This has an associated cost, described by the first term of Eq.~(\ref{w}), which must be developed by the  driving sources. In the opposite situation where
$\int_0^\tau dt\;\Vec{\Lambda}\cdot d_t \vec{B}\geq 0$, the first term of Eq.~(\ref{w}) may overcome the second one, enabling the mechanism of work output. This has an associated
extra heat transfer from the hot to the cold reservoirs, which is accounted for the first term of Eq.~(\ref{q}). This operation corresponds to a \emph{heat engine}. 


\section{Geometry of the problem}
\label{sec:geom-opt}

We now elaborate on the geometrical interpretation of the quantities presented in the previous section.

First, we factorize the total duration $\tau$ in the expressions Eqs.~\eqref{w} and \eqref{q}, such to decouple the time-rescaling from the geometrical contribution to the different quantities. 
Indeed by considering an adimensional time unit $\theta$ such that
\begin{align}
    \vec{ B}(t)=\vec{ B}(\theta \tau)\;,\quad \theta\in[0,1]\;,
\end{align}
we can define, identifying from now on the adimensional time derivative $\dot{\vec  B} \equiv{\partial \vec B}/{\partial\theta}=\tau d_t \vec{B}$,
\begin{align}
\label{eq:geometric_A}
    A=& \int_0^1 d\theta\; \Vec{\Lambda}\cdot\dot{\vec B}\;,
\\
\label{eq:geometric_L}
    L^2=& \int_0^1 d\theta\; \dot{\vec B} \cdot \underline{\Lambda}\cdot\dot{\vec B}\;,
\\
\label{eq:geometric_k}
    \langle\kappa\rangle=& \int_0^1 d\theta\; \kappa\;.
\end{align}

Accordingly, Eqs. \eqref{w} and \eqref{q} can be expressed as follows,
\begin{eqnarray}
\label{eq:def_W}
    W 
    &= & \frac{\Delta T}{T}A - \frac{L^2}{\tau} \label{w1}\\
    Q 
    &= & A+\frac{\Delta T}{T}\tau\langle\kappa\rangle. \label{q1}
\end{eqnarray}

The names $A$ and $L^2$ are related the geometrical meaning of the quantities above, as we discuss below. The representation of Eq.~(\ref{eq:geometric_A}) highlights the fact  that $A$ corresponds to a Berry-type phase in the parameter space as discussed in Ref. \cite{adiageo}. Notice, that, in order to have a non-vanishing value of $A$, at least two time-dependent parameters are necessary. 
This is basically the same argument widely discussed in the literature of adiabatic charge pumping
~\cite{brouwer1998oct,avron2001nov,moskalets2002nov,arrachea2006}.
In addition, it is necessary to break some symmetries in the system to have a finite value of this closed integral \cite{adiageo}, as discussed below.

Given that $\vec B(\theta)$ represents a closed trajectory in space, we can use Stokes' theorem --in a three-dimensional space or its corresponding generalization in higher dimensions-- to re-express the line-integral defining $A$
\begin{align}
\label{eq:definition:A}
    A=\int_{\partial\Sigma} \vec{\Lambda} \cdot d\vec{B}
    =\int_\Sigma (\vec{\nabla}_B\wedge \vec{\Lambda} )\cdot d\vec{\Sigma}\;,
\end{align}
where $\Sigma$ is a surface in the $\vec B$ space, with boundary $\partial\Sigma$ coinciding with the control trajectory. In the case of having 4 or more parameters, Eq.~\eqref{eq:definition:A} should be replaced by the Generalized Stokes' Theorem applied to differential forms in the appropriate dimension~\footnote{Identifying $\vec \Lambda \cdot d\vec B$ with a 1-form $\omega$ over $\mathbf{R}^n$, we can express Eq.~\eqref{eq:definition:A} as $A = \int_{\partial \Sigma} \omega = \int_{\Sigma} d\omega$, where $d\omega$ is the exterior derivative of $\omega$.}.
In this representation, $A$ is the flux of the vector $\vec{\nabla}_B\wedge \vec{\Lambda}$ through the area enclosed by the control trajectory, and can be also interpreted as the integral over this area weighted by the Berry curvature \cite{berry1984quantal}. 
We can therefore \emph{think of $A$ as the area of the surface defined by the control trajectory} (with local weight depending on the Berry curvature). Note that this geometrical translation clarifies as well that $A$ depends $\emph{only}$ on the geometry of the trajectory $\vec B(\theta)$: that is, not only $A$ is independent of $\tau$, but it is also invariant under any reparametrization $\theta'(\theta)$ which might change the local speed and time spent on different points of the trajectory.

Concerning $L^2$, \emph{it can be interpreted as a length squared of the control trajectory} $\vec{B}(\theta)$, as it is clear from~\eqref{eq:geometric_L} that it represents the integral of a quadratic form that defines a metric in the $\vec B$ space. At the same time, given the presence of two time derivatives, $L^2$ can depend in general on reparametrizations $\theta'(\theta)$. However, $L^2$ represents losses due to dissipation in the driving -- see Eq.~\eqref{w} -- and we are therefore interested in its minimum value, which can be obtained through a Cauchy-Schwarz inequality
\begin{align}
\label{eq:definition:L}
    L^2\geq \left(\int_0^1 d\theta\;\sqrt{\dot{\vec B} \cdot \underline{\Lambda}\cdot\dot{\vec B}}\right)^2=\left(\int_{\partial\Sigma}\sqrt{d\vec{B}\cdot \underline{\Lambda}\cdot d\vec{B}}\right)^2\equiv\mathcal{L}^2\;.
\end{align}
The lower bound $\mathcal{L}$ is fully geometric (it depends solely on $\partial\Sigma$) and it is always achievable by choosing the time-parametrization $\theta'$ such that $\dot{\vec B} \cdot \underline{\Lambda}\cdot\dot{\vec B}$ is constant. $\mathcal{L}$ is a natural extension of the standard thermodynamic length \cite{weinhold1975sep,salamon1980jul,salamon1983sep,nulton1985jul,schlogl1985dec,andresen1996nov,diosi1996dec,crooks2007sep,sivak2012may,deffner2013feb,bonanca2014jun,scandi2019oct} to non-equilibrium set-ups where the WS is simultaneously interacting with several baths. 

Finally, it is apparent that $\langle\kappa\rangle$ Eq.\eqref{eq:geometric_k} represents the simple average of a scalar number (the heat conductance) along the trajectory. In general it clearly also depends on reparametrizations of the adimensional time $\theta'(\theta)$, as the average can be arbitrarily close to the maximum value $\kappa_{\max}$ of the trajectoy, in case $\theta'$ is such to spend almost all the time close to $\kappa_{\max}$. Similarly $\langle\kappa\rangle$ can be arbitrarily close to the minimum value along the trajectory $\kappa_{\min}$.

\section{Performance of the machine and time-optimization}
\label{sec:timeopt}

In this section we discuss the different operation modes of the thermal machine, and introduce the relevant figures of merit for its characterization. 

\subsection{Heat engine}
\label{sec:heat_eng}
\begin{figure}
    \centering
    \includegraphics[width=0.5\textwidth]{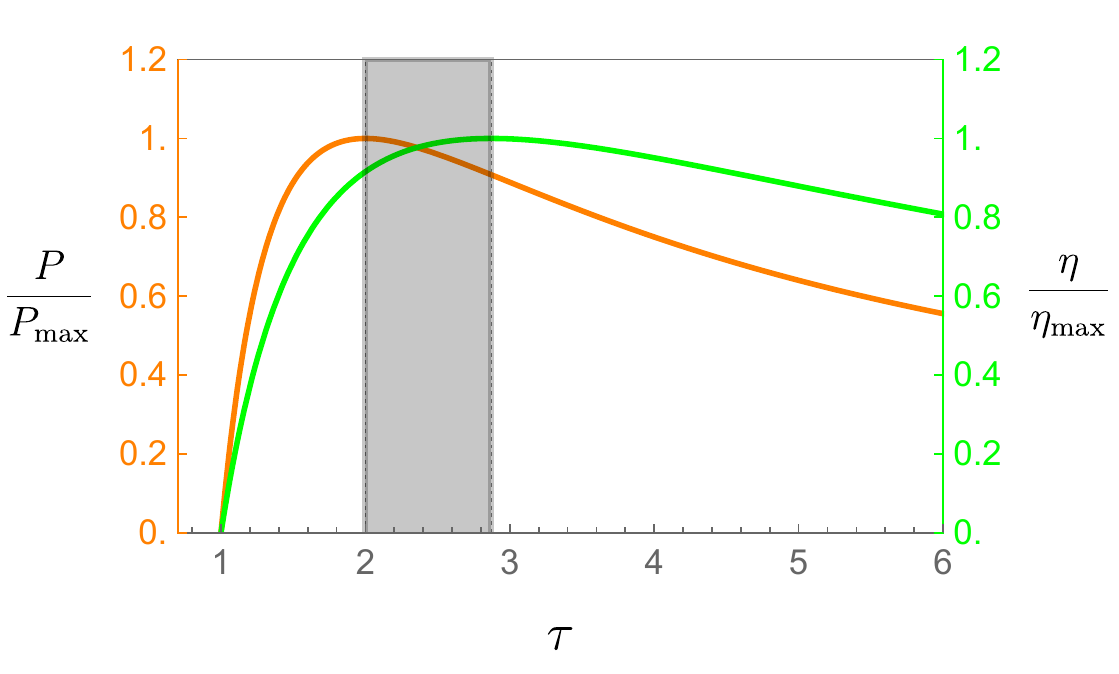}
    \caption{\underline{Engine mode: Power and efficiency vs. cycle duration.} \\
    The optimal operating region is the gray interval between the two dashed lines: indeed for any point outside the region, there is a point inside with both larger efficiency and larger power.\\
    In the limit of big heat leaks $\langle\kappa\rangle$ the corresponding heat leaks timescale $\tau_\kappa$ \eqref{eq:timescales} is small, and the difference between $\tau_P$ and $\tau_\eta$ \eqref{eq:opt_taus} shrinks. That is, when the heat leak is the dominant loss, power and efficiency maximization tend to coincide, as one could expect (this can be verified by direct inspection of \eqref{eq:pow_def} and \eqref{eq:eta_def}); the corresponding maximum efficiency is also small in this limit. \\
    In the opposite limit of no leaks $\langle \kappa\rangle\rightarrow 0$, $\tau_\kappa$ tends to infinite, and we recover the standard scenario in which power is maximized for a finite time, while the efficiency is maximum for $\tau\rightarrow\infty$, where it tends to the Carnot efficiency, as the dominant loss is due to finite-time dissipation. For finite values of $\langle \kappa\rangle$, the scenario is intermediate. In the plot $\tau_D=1$ and $\tau_\kappa=2.5$.
    }
    \label{fig:pow-eff_engine}
\end{figure}

The system described in the previous sections can be used to extract work from two reservoirs with a temperature bias. This is the \emph{engine} operating mode of the system.
We write the power of the heat engine and its efficiency as
\begin{align}
\label{eq:pow_def}
P=&\frac{W}{\tau}=\frac{\Delta T}{T}\frac{A(1-\frac{\tau_D}{\tau})}{\tau}\;,\\
\label{eq:eta_def}
\eta=&\frac{W}{Q}=\eta_C
\frac{1-\frac{\tau_D}{\tau}}{1+\frac{\tau}{\tau_k}}\;,
\end{align}
where we substituted Eqs.~\eqref{w}-\eqref{q} and we defined the dissipation  and heat leak timescales
\begin{align}
\label{eq:timescales}
   \tau_D=\frac{T}{\Delta T}\frac{L^2}{A}\;,\quad \tau_\kappa= \frac{T}{\Delta T}\frac{A}{\langle\kappa\rangle}\;.
\end{align}
In the previous expressions $\eta_C=\Delta T/T$ is the Carnot efficiency.
Given the expressions above, we can optimize the duration of the cycles in order to maximize the power or the efficiency, obtaining correspondingly
\begin{align}
    \tau_{P}=2\tau_D,\quad \tau_\eta=\tau_D+\sqrt{\tau_D(\tau_D+\tau_\kappa)}\;.
    \label{eq:opt_taus}
\end{align}
We see that the duration for maximum efficiency is always larger than the duration for maximum power.
The corresponding maximum power and efficiency at maximum power are
\begin{align}
\label{eq:Pmax}
P_{\rm max}=\frac{1}{4}\frac{(\Delta T)^2}{T^2}\frac{A^2}{L^2}\;,\quad \eta_{P_{\rm max}}=
\frac{\eta_C}{2} \frac{x-1}{x+1}
\end{align}
while the maximum efficiency and power at maximum efficiency
\begin{align}
\label{eq:Etamax}
\eta_{\rm max}=
\eta_C\left(1-\frac{2}{\sqrt{x}+1}\right)\;,\quad P_{\eta_{\rm max}}=\frac{(\Delta T)^2}{T^2}\langle \kappa\rangle \frac{(\sqrt{x}-1)^2}{\sqrt{x}}\;,
\end{align}
with 
\begin{equation}\label{eq:x}
x=1+\frac{A^2}{L^2\langle\kappa\rangle}.
\end{equation}
See Fig.~\ref{fig:pow-eff_engine} for a summary and visual explanation of these results.

\subsection{Refrigerator}
\label{sec:heat_pump}
In the \emph{heat pump} or \emph{refrigerating} mode, external work is supplied  to the system to extract heat from the cold bath and transfer it to the hot one. Therefore we define the cooling power $P'$ and the coefficient of performance (COP) $\eta'$ 
\begin{align}
\label{eq:P'ref}
    P'=&\frac{-Q}{\tau}=A\frac{1-\frac{\tau}{|\tau_k|}}{\tau}\;,\\
\label{eq:eta'ref}
    \eta'=&\frac{Q}{W}=\eta'_C 
    \frac{1-\frac{\tau}{|\tau_k|}}{1+\frac{|\tau_D|}{\tau}}\;,
\end{align}
where $\eta'_C =T/\Delta T$ is the Carnot COP.
The difference with the engine operating mode is that in this case both $Q$ and $W$ are negative (heat is transferred against the thermal bias and work is performed \emph{on} the system).
We have therefore $A<0$ which implies $\tau_\kappa<0$ and $\tau_D<0$ are formally negative as well (which is the reason of the absolute values in the equations). 
By direct inspection of~\eqref{eq:P'ref} we see that the maximum power of such mode is unbounded, as in the limit $\tau\rightarrow 0$ the power tends to infinity. 
The slow-driving approximation $\tau_{\rm rel}/\tau \ll 1$ prevents us from analyzing the limit of arbitrary small $\tau$ and a reliable analysis of the cooling power requires a description beyond linear response~\cite{hajiloo2020detailed,mateos2021thermoelectric}. Thus, we focus only on maximizing the efficiency of this operation, for which we get
\begin{align}
 \tau_{\eta'}&=\tau_{\eta}=\sqrt{\tau_D(\tau_D+\tau_\kappa)}-|\tau_D|\;,
\end{align}
\begin{align}
    \eta'_{\rm max}=\frac{T}{\Delta T}\left(1-\frac{2}{\sqrt{x}+1}\right)\;,
\end{align}
\begin{align}
    P'_{\eta'_{\rm max}}=\frac{\Delta T}{T}\langle \kappa\rangle \sqrt{x}\;,
\end{align}
with $x$ defined as in Eq.~(\ref{eq:x}).

\section{Full optimization and the isoperimetric problem}
\label{sec:results}
In the previous section we showed how to choose the optimal duration for cycles of two kinds of thermal machines, and we derived formal expressions for the resulting powers and efficiencies. The resulting figures of merit still depend on the particular trajectory chosen for the cycle. Finding the fully-optimal solution is nontrivial, but we show in the following how the geometrical picture of the thermodynamics introduced in Sections~\ref{sec:geom-opt} and \ref{sec:timeopt}, helps in finding the most advantageous control trajectories to be exerted on the machine.

An interesting question in the present problem is whether we can find a protocol that maximizes the output power of the system. We have shown in Section~\ref{sec:geom-opt} that, given a parametrization $\vec B(\theta\tau)$ defined over $\partial \Sigma$ in the parameter space, we can compute the duration $\tau$ that upper bounds the power for that protocol. The result is expressed in Eq.~\eqref{eq:Pmax}.
Besides, we know from the definition in Eq.~\eqref{eq:definition:A} that the value of $A^2$ does not depend on reparametrizations $\theta'$, while the value of $L^2$ can be lower-bounded by $\mathcal{L}^2$ according to Eq.~\eqref{eq:definition:L}.
With all these considerations, we find that the maximum power developed by a protocol moving along a curve $\partial \Sigma$ is expressed by
\begin{align}
\label{eq:PmaxC}
    P_{{\rm max}} (\partial \Sigma)=
    \frac{1}{4}\frac{(\Delta T)^2}{T^2}
    \frac{A^2}{\mathcal{L}^2}.
\end{align}

Eq.~\eqref{eq:PmaxC} tells us that the problem of finding the maximum output power of the system is equivalent to the problem of maximizing the term~$A^2 / \mathcal{L}^2$ over the set of all closed curves~$\partial \Sigma$ in the parameter space (known as isoperimetric or Cheeger problem \cite{ros2001isoperimetric,parini2011introduction,leonardi2015overview}).
The optimization of this geometrical quantity is not a simple task in general, since one must choose a test curve $\partial \Sigma$ that maximizes~$A^2$, while keeping~$\mathcal{L}^2$ small, being those quantities  nontrivial functions of~$ \partial \Sigma$ when the corresponding metrics are not flat \cite{howards1999isoperimetric,morgan2005manifolds,rosales2008isoperimetric,carroll2008isoperimetric}.

For what concerns the efficiencies, $\eta_{\rm max}$, $\eta'_{\rm max}$, $\eta_{P_{\rm max}}$ are all increasing functions of the same parameter $A^2/(L^2\langle\kappa\rangle)$. Like in Eq.~\eqref{eq:definition:L} the denominator can be lower bounded with a Cauchy-Schwarz inequality
\begin{multline}
\label{eq:CS_eff}
    L^2\langle \kappa\rangle=\left(\int_0^1 d\theta\; \dot{\vec{B}} \cdot \underline{\Lambda}\cdot\dot{\vec{B}}\right)\left( \int_0^1 d\theta\; \kappa\right)\\
    \geq \left(\int_0^1 d\theta\;\sqrt{\kappa} \sqrt{\dot{\vec{B}} \cdot \underline{\Lambda}\cdot \dot{\vec{B}}}\right)^2\;.
\end{multline}
In complete analogy to Eq.~\eqref{eq:definition:L}, the bound can be always saturated, by choosing  a reparametrization $\theta'(\theta)$ such that $\dot{\vec{B}} \cdot \underline{\Lambda}\cdot\dot{\vec{B}}/\kappa$ is constant in time, and can be interpreted again as a length defined by an underlying metric
\begin{align}
      \left(\int_{\partial \Sigma}\sqrt{d\vec{B} \cdot \underline{\Lambda_\kappa} \cdot d\vec{B}}\right)^2\equiv \mathcal{L}^2_\kappa\;,
\qquad
\underline{\Lambda_\kappa}=\underline{\Lambda}\kappa\;.
 \label{eq:LambdaK_def}
\end{align}
The length $\mathcal{L}_\kappa$ is fully geometric, i.e. it depends only on the set of points defined by the trajectory $\partial\Sigma$, and the maximization of $\eta_{\rm max}$, $\eta'_{\rm max}$, $\eta_{P_{\rm max}}$ is also mapped to an isoperimetric problem
\begin{align}
\label{eq:etamaxC}
    \max \frac{A^2}{L^2\langle\kappa\rangle}=\max_{\partial\Sigma} \frac{A^2}{\mathcal{L}^2_\kappa}\;.
\end{align}
The geometric expressions \eqref{eq:PmaxC} and \eqref{eq:etamaxC}, which map the thermodynamic optimization to an isoperimetric (Cheeger) problem,   are the main results of this paper.

\section{A qubit thermal machine}\label{sec:qubit}
We will exemplify these results for the specific case of a driven qubit, in which case, the Hamiltonian for the working substance entering Eq.~(\ref{hamtot}) is ${\cal H}_{\rm WS}(t)={\cal H}_{\rm qb}(t)$, where
\begin{equation}\label{qs}
{\cal H}_{\rm qb}(t)= \vec{ B}(t) \cdot \hat{\vec{\sigma}}
\end{equation}
with $\hat{\vec{\sigma}}=(\hat{\sigma}_z,\hat{\sigma}_x)$ being  the Pauli matrices and $\vec{ B}(t) \equiv \left(B_z(t), B_{x}(t) \right)$,
being periodic with period $\tau$. 

As already highlighted in  Section~\ref{sec:setup}, a key ingredient  to have  the  heat--work mechanism  in the linear response regime, is some protocol leading to $A \neq 0$. We recall that this quantity represents also the net pumped heat as a consequence of the time-dependent driving. 
In linear response, $A$ depends on response functions that are evaluated with the two reservoirs at the same temperature $T$ (see \cite{ludovico2016feb,adiageo} and Appendix \ref{app-lin-res}). When the two reservoirs are equally coupled, 
any protocol implemented via changing $\vec{B}$ generates the same energy flow between them and the qubit. This  prevents a net energy transfer between the reservoirs  and $A=0$. Therefore, it is necessary to introduce some asymmetry in the coupling between the qubit and the reservoirs in order to have $A \neq 0$.
For this reason, we consider the Hamiltonian describing the coupling to the reservoirs introduced in  Eq.~\eqref{qcont} with $\hat{\pi}_{\rm h} \equiv \hat{\sigma}_{x}, ~\hat{\pi}_{\rm c}=\hat{\sigma}_z$, which breaks the c $\leftrightarrow$ h  symmetry in the absence of a temperature bias. Any other combination of Pauli matrices with
$\hat{\sigma}_{\rm h}\neq \hat{\sigma}_{\rm c}$ would lead to similar results. As mentioned before, the other crucial ingredient is a protocol depending on at least two parameters, which is necessary to define a non-trivial surface $\Sigma$.  In our case, we consider just two parameters: $B_z(t)$ and $B_x(t)$.

We solve the problem in the limit of weak coupling between the WS and the reservoirs by deriving the adiabatic master equation by means of  the non-equilibrium Green's function formalism at second order of perturbation theory in $V_{\alpha}$ as explained in Ref.~\cite{masterkeldysh}. Details are shown in Appendices ~\ref{sec:linear-response} and ~\ref{appendix:sameCouplingStrength}.
In the specific calculations discussed below, we considered the simplest case, where the two reservoirs have the same spectral density, $\Gamma_{\rm c}(\epsilon)=\Gamma_{\rm h}(\epsilon)=
\Gamma(\epsilon)=\bar \Gamma \epsilon e^{-\epsilon/\epsilon_C}$ for $\epsilon \geq 0$. 

\subsection{Adiabatic linear-response coefficients}
The adiabatic linear response matrix $\underline{\Lambda}$ is originally expressed as a function of the coordinates $(B_z,B_x)$. This matrix is positive defined and symmetric. 
When diagonalized it is found that the eigenvectors, $|r\rangle=\left(\sin(\phi), \cos(\phi)\right)^T$, $|\phi\rangle=\left(\cos(\phi), -\sin(\phi)\right)^T$ 
correspond to radial and tangential directions and thus $\underline\Lambda$ can be expressed as follows,
\begin{equation}\label{eq:underlam}
\underline{\Lambda} =
\lambda_{r}|r\rangle \langle r|+
\lambda_{\phi}|\phi\rangle \langle \phi|,
\end{equation}
with $\lambda_{r}, \lambda_{\phi}\geq 0$.
This suggests that it is natural to implement the following change of coordinates $B_z=B_r\cos\phi, \; B_x=B_r\sin\phi$. We get
\begin{equation}
\dot{\vec B} \cdot \underline{\Lambda} \cdot \dot{\vec B} \equiv {\lambda_{r}} {\dot B}_r^2 + 
\lambda_{\phi} B_r^2 {\dot \phi}^2
\equiv
\lambda_{r} |\dot{\vec B}_r|^2
+ \lambda_{\phi} |\dot{\vec B}_\phi|^2.
\end{equation}


 The analytical expression for the radial component reads,
\begin{equation}
\label{eq:lambda_diss_radial-t}
\lambda_{r}(\vec B)=
\frac{\hbar \beta \sinh (\beta  B_r) }{\Gamma (2 B_r)\cosh^3{(\beta B_r)}
}
\end{equation}

and for the tangential one
 is
\begin{equation}
\label{eq:lambda_diss_polar-t}
\lambda_{\phi}(\vec B)=
\frac{\hbar \Gamma (2 B_r)}{4 B_r^3}
,
\end{equation}
being $\beta=1/k_B T$.
The first component is associated to changes in the energy gap between the two states of the qubit, while the second one leaves the spectrum unchanged but introduces a rotation of the eigenstate basis. 

Regarding the other coefficients, the components of the vector $\vec{\Lambda}({\vec B})=(\Lambda_z, \Lambda_x) = \Lambda_r \langle r| + \Lambda_\phi \langle \phi|$ 
read
\begin{equation}
    \Lambda_{r}(\vec B) = \frac{\beta B_r \sin ^2(\phi)  }{\cosh^2 (\beta B_r) },\;\;\;
    \Lambda_{\phi}(\vec B) = 0\;,
\end{equation}
while the parametric thermal conductance is
\begin{equation}
\label{eq:qubit_kappa}
    \kappa(\vec B) = \frac{\beta B_r^2 \sin^2 (2 \phi ) \Gamma (2 B_r)}{\sinh{(2\beta B_r)}\hbar}.
\end{equation}

\subsection{Geometrical quantities and bound for the heat--work conversion}
\label{sec:results-C}
Given the above coefficients we can now calculate all 
 the relevant geometrical quantities for the characterization of the machine, namely, $A$, $L$ and $\langle\kappa\rangle$ defined respectively in Eqs. \eqref{eq:geometric_A}, \eqref{eq:geometric_L} and \eqref{eq:geometric_k}.

\begin{figure}
    \centering
    \includegraphics[width=0.5\textwidth]{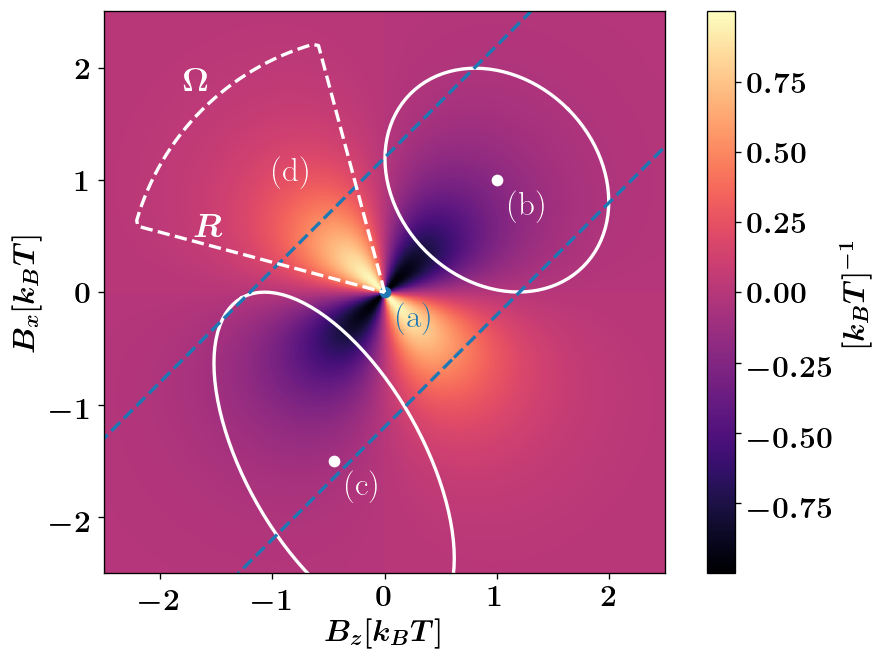}
    \caption{The Berry-type curvature $\left[\vec\nabla_B \wedge \vec\Lambda(B)\right]_y$. The integration of this quantity over the area enclosed by the control trajectory defines the $A$ as in Eq. (\ref{eq:definition:A}). Parameters are $\epsilon_C=120 k_B T$ and $\bar\Gamma=0.2$. Curves (a), (b) and (c) 
    are heuristically searched protocols of elliptic shape, centered in $(0,0)$, $(1,1)$ and $(-1.5, -0.45)$ respectively, that maximize the value $A^2/\mathcal{L}^2$ (see Sec. \ref{subsec:maximum_power}). Curve (d) is a protocol with the shape of a circular sector
    centered at $(0,0)$, with radius $R$ and spanning an angle $\Omega$ symmetrically with respect to the quadrant's bisector.
    }
    \label{fig:curl_lambda}
\end{figure}
As already mentioned when  the representation of Eq.~\eqref{eq:definition:A} was introduced, the net pumped heat quantified by $A$ is simply the value of the Berry curvature integrated over the area of the $(B_z,B_x)$ plane enclosed by a particular protocol. 
 The  Berry curvature as a function of $(B_z,B_x)$ is shown in Fig. \ref{fig:curl_lambda}.
Because of the nature of the setup,  this quantity changes sign at $B_x=0$ and $B_z=0$. Therefore, 
protocols with constant $B_r$ lead to $A=0$. For any protocol, the sign can be simply switched by changing the circulation of the boundary curve, hence switching  the operation from heat engine to refrigerator or viceversa.

It is also easy to visualize in Fig. (\ref{fig:curl_lambda}), that protocols enclosing a large portion of the dark blue  or bright yellow areas lead to a large value of  $|A|$.
Focusing on simple curves that do not cross themselves we consider a circular-sector trajectory like the curve $(d)$ depicted in Fig.~(\ref{fig:curl_lambda}), characterized by a radius $R$ and an aperture angle $\Omega$ symmetric with respect to the the quadrant's bisector. 
It is clear from the figure that the protocol leading to the maximum achievable value of $|A|$ in the present setup corresponds to a trajectory fully enclosing a quadrant.
Such a trajectory is, for instance, the special case of the circular-sector trajectory with $\Omega=\pi/2$ that:
i) starts at the origin and goes to infinity along the $B_x$ axis,
ii) rotates $\pi/2$ counterclockwise and aligns in the $B_z$ axis,
iii) returns to the origin along the $B_z$ axis.

This limiting protocol corresponds to a quasistatic Carnot cycle and the resulting value of $A$ is
\begin{equation}\label{landauer}
    A_{\rm lim}=\int_{\text{quadrant}} (\vec{\nabla}_B\wedge \vec{\Lambda} )\cdot d\hat{y}=\pm k_B T \log(2),
\end{equation}
where the signs are determined by the enclosed quadrant and the circulation considered. Notice that, according to Eq.~(\ref{q}),  this corresponds to the extreme values for the energy that could be transported 
between the two reservoirs at the same temperature $T$ through the qubit, and coincides with the famous bound obtained by Landauer's argument~\cite{landauer61} according to which the change of Shannon entropy in the process of erasing the information encoded in a bit  is $\pm \log(2)$.  In the present case, it is associated to the transfer of the same amount of entropy  between the reservoirs (a similar result was found quantum dots \cite{janine}). At finite $\Delta T$, according to Eq.~(\ref{w}) this quantity also sets the maximum value of the work that can be extracted in the heat-engine operational mode (for $A_{\rm lim}>0$) in the limit of
 vanishing dissipation. This  result is, respectively,
\begin{equation}
\label{eq:limitW}
    W_{\rm lim}=  k_B (T_{\rm h}- T_{\rm c}) \log(2)= A_{\rm lim} \; \eta_C.
\end{equation}


\begin{figure}
    \centering
    \includegraphics[width=0.45\textwidth]{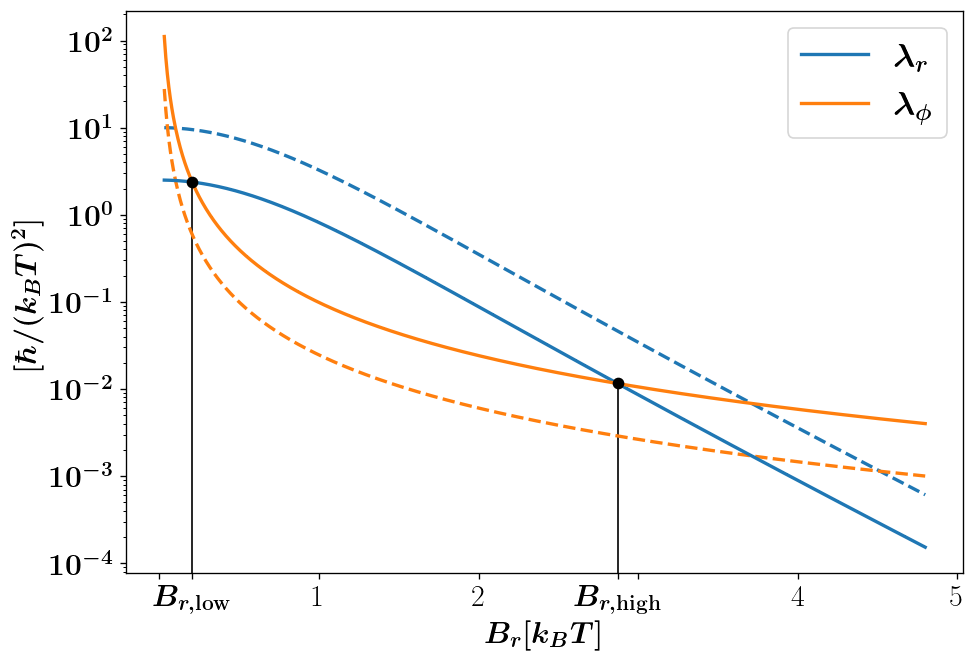}
    \caption{Positive eigenvalues of $\underline{\Lambda}$ -see Eq. \eqref{eq:underlam}- as a function of $|\vec B|=B_r$. Parameters are $\epsilon_C=120 k_B T$ and $\bar\Gamma=0.2$ (solid lines), $\bar\Gamma=0.05$ (dashed lines). Note for $\bar \Gamma=0.2$ (solid lines) that most of the relevant region of Fig. \ref{fig:curl_lambda} lies inside the interval $(B_{r,\rm low}, B_{r, \rm high})$ where the radial dissipation is about one order of magnitude bigger than the polar dissipation.}
    \label{fig:dissipation}
\end{figure}

We now turn to analyze $L^2$, which assesses
the dissipated energy for a particular protocol.
This  quantity is  determined by $\underline{\Lambda}$ given by Eq.~(\ref{eq:geometric_L}). For the qubit, 
this matrix can be decomposed in two contributions, as expressed in  Eq.~(\ref{eq:underlam}) which are associated to   the dissipation of energy  originated in the radial and polar changes of $\vec{B}$.

We see from the analytical expressions of Eqs. (\ref{eq:lambda_diss_radial-t}) and (\ref{eq:lambda_diss_polar-t}) that $\underline{\Lambda}$ is symmetric along the polar axis, i.e. it only depends on $B_r$. This is illustrated in the upper panel of Fig. \ref{fig:dissipation_change} of Appendix~\ref{appendix:sameCouplingStrength}. In Fig. \ref{fig:dissipation} we show the dependence of the coefficients $\lambda_{r}$ and $\lambda_{\phi}$ on $B_r$ for two different values of the $\bar \Gamma$ parameter. For some values of $\bar\Gamma$ we find an interval $(B_{r,{\rm low}}, B_{r,{\rm high}})$ for which the dissipation is mainly due to changes in the energy spectrum induced by finite $\dot B_r$.
The specific values  $B_{r,\rm low},\; B_{r,\rm high}$
depend on the working temperature and the coupling constant $\bar \Gamma$ between the qubit and the reservoirs. More details on the $\underline{\Lambda}$ submatrix and the dissipation structure of the qubit can be found in Appendix~\ref{appendix:sameCouplingStrength}.

The final value of $L^2$ for a protocol $\vec B(\theta \tau)$ defined over $\partial \Sigma$ in the parameter space still depends on the chosen parametrization $\theta$.
Out of all the possible  parametrizations, Eq.~\eqref{eq:definition:L} tells us that there exists a particular one for which
 $L^2=\mathcal{L}^2$.
Furthermore, this corresponds to the lower bound for $L^2$ and, importantly,  it  is  a function of $\partial \Sigma$ only (it is geometrical).

In addition, for a given $\theta$ associated to $\partial \Sigma$, we are able to obtain the optimal parametrization $\bar\theta (\theta)$ that saturates the bound, and defines the less dissipative protocol $\vec{B} (\bar{\theta}\tau)$ around $\partial \Sigma$ in time $\tau$. The new value of the velocity at a given time can be computed using \eqref{eq:definition:L}, demanding that $\dot{\vec B}(\theta \tau)
\cdot \underline{\Lambda} (\vec B)
\cdot \dot{\vec B}(\theta \tau)$ is constant at each point.
The result is
\begin{align}
\label{eq:optimal_parametrization_velocity}
   \frac{\partial \vec {B} (\bar{\theta}\tau)}{\partial \bar{\theta}}=
    \dot{\vec B}(\theta \tau)
    \sqrt{\frac{\mathcal{L}^2}
    {\dot{\vec B}(\theta \tau)
    \cdot \underline{\Lambda} (\vec B)
    \cdot \dot{\vec B}(\theta \tau)}
    }
\end{align}
where the dot in $\dot{\vec{B}}$ is the derivative with respect to the original parametrization $\theta$. This driving ensures constant entropy production along the cycle. 

\subsection{Maximum power}
\label{subsec:maximum_power}
Although a global maximum for $P_{\rm max}(\partial\Sigma)$ in Eq.~\eqref{eq:PmaxC} is hard to find, it is still possible to design simple trajectories with useful output power and reasonable efficiency. 
We perform a numerical search of $\mbox{max}_{\partial \Sigma} A^2/\mathcal{L}^2$ using a gradient descent method, restricted to the space of elliptic trajectories centered at a given point $\vec B$. The trajectories $(a)$, $(b)$ and $(c)$ shown in Fig.~\ref{fig:curl_lambda} are examples of the resulting curves. We choose this type of curves because  elliptical trajectories are easy to implement and flexible enough to perform an extensive optimal search.
The advantage of the elliptical protocols is not obvious, taking into account that Fig.~\ref{fig:curl_lambda} suggests that
the circular-sector protocols are better than the ellipses  for maximising~$A$. However, this is not the case for~$A^2/\mathcal{L}^2$: we show in Appendix~\ref{app:pizzaProtocol} that suitable chosen ellipses can clearly outperform circular-sector protocols in terms of power output. 

Focusing on the elliptic protocols, we see that the highest values of power are achieved for test curves that avoid the region of small $|\vec B|$, where the dissipation coefficient $\lambda_{\phi}$ diverges. The curve $(a)$ centered at~$(0,0)$ is an interesting example. It maximizes $A^2$ by enclosing the two lobes in the first and third quadrant of Fig. \ref{fig:curl_lambda}, and closes the curve near infinity in order to avoid the central region of high dissipation.

\begin{figure}
    \centering
    \includegraphics[width=0.45\textwidth]{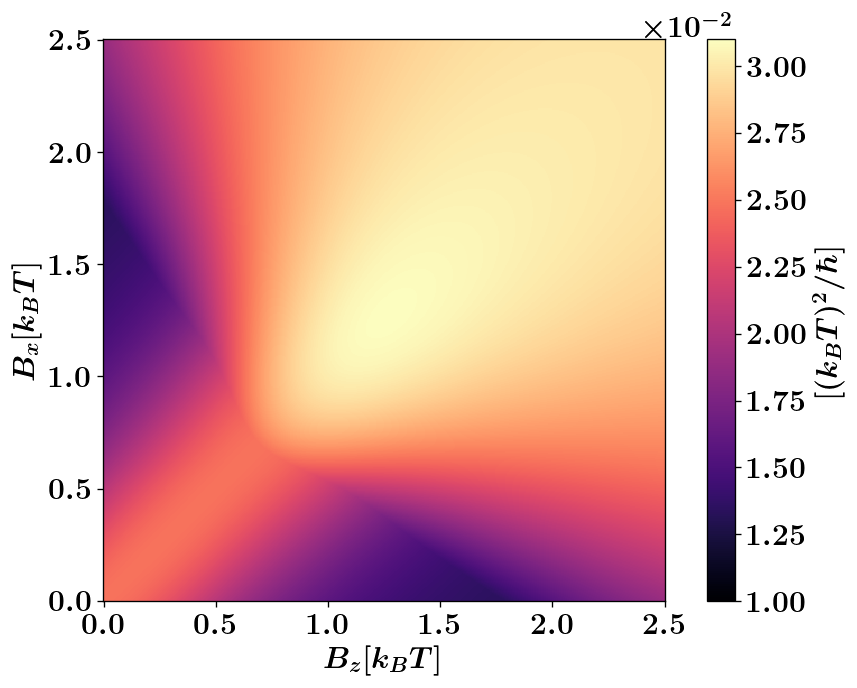}
    \caption{max $A^2/\mathcal{L}^2$ as a function of $\vec B$, for an heuristic optimization of elliptic trajectories centered at $\vec B$. Parameters are $\epsilon_C=120 k_B T$ and $\bar\Gamma=0.2$. Only positive values of $B_x$ and $B_z$ are shown, since this quantity is symmetric with respect to $B_x=0$ and $B_z=0$.} 
\label{fig:optimal_ellipses}
\end{figure}

In Fig.~\ref{fig:optimal_ellipses} we depict the value of $\mbox{max}_{\partial \Sigma} A^2/\mathcal{L}^2$ found by the mentioned heuristic method, as a function of the (fixed) central point of the ellipse.
 We distinguish two different regimes leading to the optimal power, as a consequence of the crossover between the two mechanisms of dissipation discussed in the context of Fig. \ref{fig:dissipation}. For small $B_r$, where the less dissipative protocol is radial, the optimal trajectories are like the case (a) in Fig. \ref{fig:curl_lambda}, while in the opposite limit where $B_r$ is large, the optimal protocols are like the ones indicated with (b) and (c) in that Fig.

\subsection{Maximum efficiency}
\label{sec:max_efficiency}

\begin{figure}
    \centering
    \includegraphics[width=0.5\textwidth]{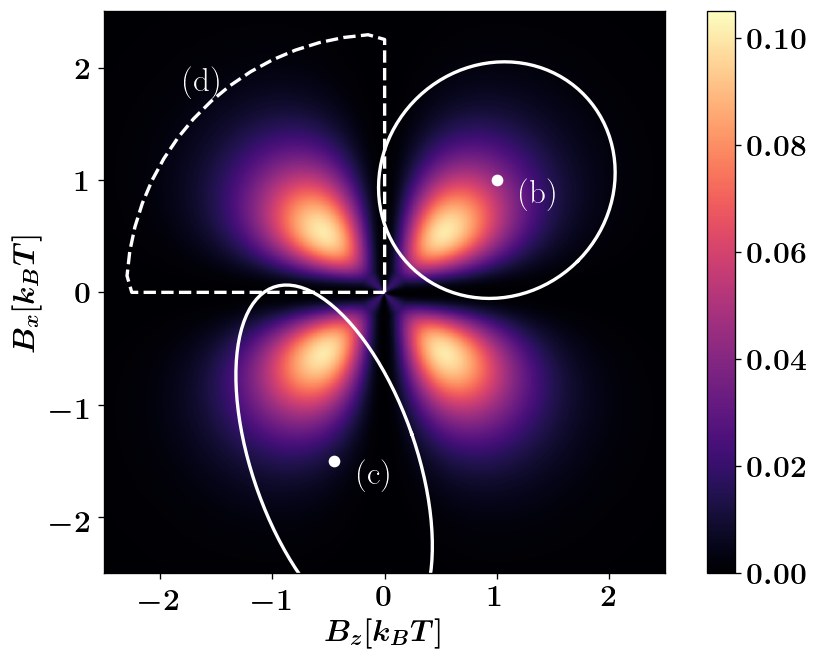}
    \caption{Maximum eigenvalue of $\underline{\Lambda}_{\kappa}$,
    indicating the losses due to the combined effect of dissipation and thermal conduction. Parameters are $\epsilon_C=120 k_B T$ and $\bar\Gamma=0.2$. Curves (b) and (c) are heuristically searched protocols of elliptic shape, centered at $(1,1)$ and $(-1.5, -0.45)$ respectively, that maximize the value $A^2/\mathcal{L}_\kappa^2$. Curve (d) (circular sector) is a quarter of circumference centered at $(0,0)$, joined by two radial lines of length $R$ along the axis.}
    \label{fig:lambdaK}
\end{figure}

Following the same philosophy of  the analysis of $L^2$ presented in Fig.~\ref{fig:dissipation}, we plot in  Fig.~\ref{fig:lambdaK} the maximum eigenvalue of 
$\underline{\Lambda}_{\kappa}$ (defined in Eq.~\eqref{eq:LambdaK_def}) 
in order to visualize the value of the thermal losses when the system evolves in the direction of maximum dissipation. Note that since $\kappa(\vec B)$ is a scalar, hence, an equivalent decomposition to Eq.~\eqref{eq:underlam} can be done for $\Lambda_\kappa^2$ as follows:
%
\begin{equation}
\label{eq:underlamKappa}
\underline{\Lambda_k} =
{\lambda_k}_{r}|r\rangle \langle r|+
{\lambda_k}_{\phi}|\phi\rangle \langle \phi|.
\end{equation}
Furthermore the analysis presented for $L^2$ in Section~\ref{sec:results-C}, and particularly the results shown in Fig.~\ref{fig:dissipation} and Appendix \ref{appendix:sameCouplingStrength} still hold for the eigenvalues and eigenvectors of $\underline{\Lambda}_\kappa$.

In the case of the efficiency, an optimal solution is trivially found by looking at Fig.~\ref{fig:lambdaK} and considering again the circular-sector curve $(d)$ with $\Omega=\pi/2$. From \eqref{eq:qubit_kappa} we see that along the $B_x$ and $B_z$ axis we have $\kappa=0$, because in those regions the system is coupled to only one of the reservoirs. 
It is clear from Eq.~\eqref{eq:LambdaK_def} that for the limiting circular-sector protocol with $R \rightarrow \infty$, enclosing the full quadrant and leading to Eq.~(\ref{landauer}) we have $\left< \kappa \right>=0$, which implies $x=\infty$ in Eq.~\eqref{eq:Etamax}, hence $\eta_{\rm max}=\eta_C$. In fact, as already mentioned, this protocol is an equilibrium Carnot cycle for the qubit, where the changes along the axis are the isothermal compression and expansion. More details of the efficiency of this protocol can be found in Appendix \ref{app:pizzaProtocol}.

In addition to this particular solution of special interest, we illustrate the usefulness of the method in a more generic way. The strong equivalence between the geometrical quantities $A^2/\mathcal{L}^2$ and $A^2/\mathcal{L}_\kappa^2$ allows us to replicate the analysis done in the previous subsection in a straightforward manner. Once again, for a given a trajectory the value of $A^2$ is computed from Eq.~\eqref{eq:definition:A} while the lower bound for $L^2\left<\kappa\right>$ and the corresponding optimal parametrization is given by~\eqref{eq:LambdaK_def} in complete analogy with Eqs.~\eqref{eq:definition:L} and~\eqref{eq:optimal_parametrization_velocity} from the maximum power analysis.
 
\begin{figure}
    \centering
    \includegraphics[width=0.5\textwidth]{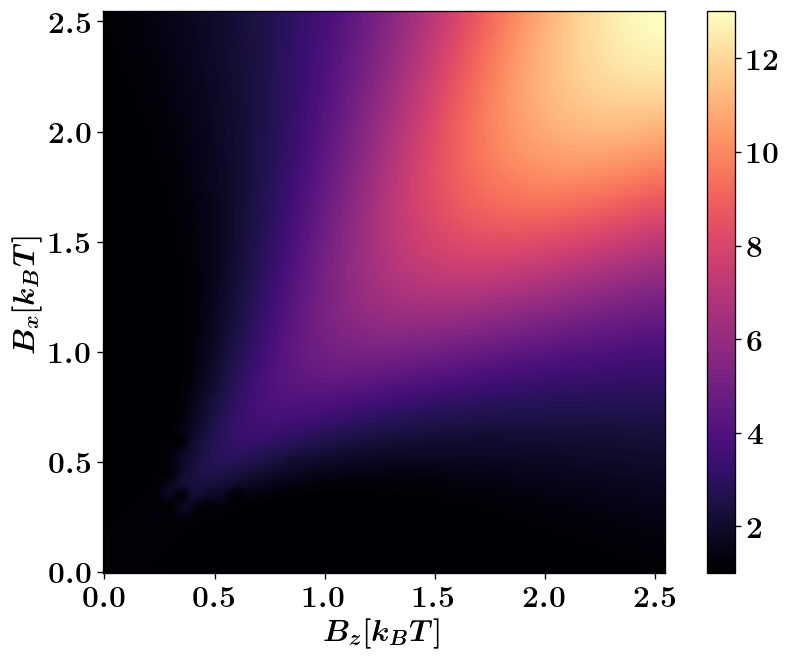}
    \caption{Maximum $A^2/\mathcal{L}_\kappa^2$ as a function of $\vec B$ for a heuristic optimization of  elliptic trajectories centered at $\vec B$. Parameters are $\epsilon_C=120 k_B T$ and $\bar\Gamma=0.2$.}
\label{fig:optimal_ellipses_lambdaK}
\end{figure}

We perform the numerical search of $\max_{\partial \Sigma}\lbrace A^2/\mathcal{L}_\kappa^2 \rbrace(\vec B)$ again for the special case of closed elliptic curves centered at a given point $\vec B$. The computed result is presented in Fig.~\ref{fig:optimal_ellipses_lambdaK}.
In Fig.~\ref{fig:lambdaK} we also show some of these trajectories, centered at the points $\vec B= \lbrace (1,1),(-1.5,-0.45) \rbrace$. Note that, while some differences can be spotted between these trajectories and the ones shown in Fig.~\ref{fig:curl_lambda}, the qualitative intuition is that efficient protocols are the ones with big $A^2$.

\subsection{The impact of optimizing the driving speed} 
The aim of this section is to gather further insight on the effect of selecting the optimal protocol, regarding  the trajectory $\partial \Sigma$ and the optimal speed for the circulation on the resulting power and efficiency of a heat engine. 

We consider an elliptical protocol for which we can define a ``trivial'' circulation with constant angular velocity. 
For the case of the power,  we compare the results of such trivial circulation with the one corresponding to 
 the optimal velocity as defined in
Eq. (\ref{eq:optimal_parametrization_velocity}). For the case of the efficiency we compare the trivial circulation with 
 with the one corresponding to the optimal velocity,
as defined in Eq. (\ref{eq:optimal_parametrization_velocity}) with the replacements ${\cal L} \rightarrow {\cal L}_{\kappa}$
and $\underline{\Lambda} \rightarrow \underline{\Lambda}_{\kappa}$.

\begin{figure}
    \centering
    \includegraphics[width=0.5\textwidth]{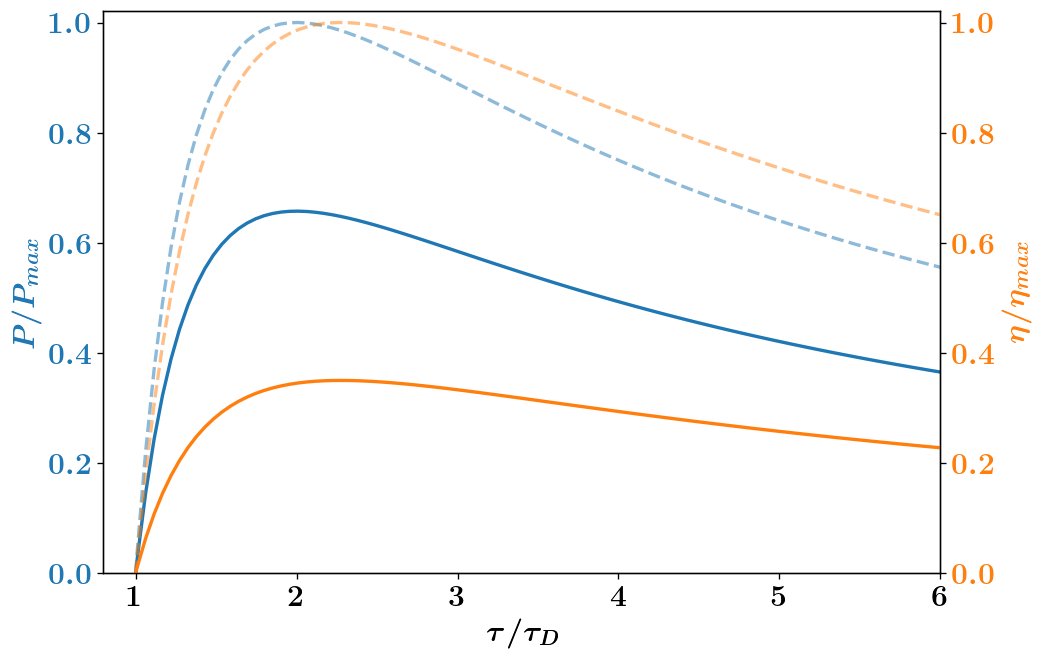}
    \caption{Power (blue) and efficiency (orange) for curve (b) of Fig. \ref{fig:lambdaK} as a function of the cycle duration $\tau$. Solid lines: circulating around the curve at constant angular velocity. Dashed lines: Using the optimal velocities given by Eqs. \eqref{eq:definition:L} (for power) and \eqref{eq:LambdaK_def} (for efficiency).}
\label{fig:curve_b_performance}
\end{figure}

For sake of concreteness we focus on $\partial \Sigma$ given by the protocol (b) of Fig.~\ref{fig:curl_lambda}. Results are shown in Fig.~\ref{fig:curve_b_performance}, where we show the power and efficiency of the machine as a function of the cycle total duration. Plots in solid and dashed lines correspond, respectively, to the protocols with constant angular velocity and optimal velocity. 
 We note from this figure that the optimized parametrization is around two times bigger in power, and around four times more efficient, with respect to the trivial parametrization of the ellipse circulated at a constant angular velocity.
Dashed lines in Fig.~\ref{fig:curve_b_performance} are akin to those of Fig.~\ref{fig:pow-eff_engine} where power values are normalized to $P_{\rm max}$ and efficiencies to $\eta_{\rm max}$, and summarizes the performance of the machine.

\subsection{Estimates for the performance}
To finalize the analysis of the qubit heat engine, it is interesting to analyze concrete values characterizing its performance.
As before, we focus on the protocol (b) of Fig. \ref{fig:curl_lambda}, 
for which we have
\begin{equation}\label{al-b}
    A^2 = 0.233 k_B^2 T^2 \quad
     \mathcal{L}^2 = 7.71 \hbar.
\end{equation}
For these values, we find using Eq.~\eqref{eq:Pmax}:
\begin{align*}
P_{\rm max}= \left( 1.364 \times 10^{-2} \frac{pW}{K^2}\right) (\Delta T)^2,
\end{align*}
which for a working temperature of $T=100mK$ and a temperature bias  corresponding to $\Delta T=0.05 T$, as in previous Figures, gives
$P_{\rm max}=  
0.341 aW $
with efficiency $\eta_{P_{max}} = 0.23\eta_C$. The total time $\tau_P$ for maximum power output per cycle is computed through Eq.~\eqref{eq:opt_taus}:
\begin{align*}
    \tau_P = 2 \tau_D = 48.8 ns
\end{align*}
which corresponds to an operation frequency in the order of $0.1GHz$. 

It is interesting to compare the value obtained for the maximum power in the protocol under consideration with the power associated to the limiting value for the work given by Eq. (\ref{landauer}). Such limiting power can be obtained by 
replacing $\tau_P$ of Eq. \eqref{eq:timescales} in Eq. \eqref{eq:def_W}, where we see that at finite time the net work done by the machine operating at maximum power is $W_{P_{\rm max}}= A \;\eta_C/2$. Taking into account that for the heat engine
$A \leq \log(2) k_B T$ -- see Eq. (\ref{landauer})-- we conclude that the bound for the maximum operating power
in a cycle of duration $\tau_P$
is $P_{\rm lim}=\log(2) \; k_B T \;\eta_C/(2\tau_P)$. For the case of the protocol (b), given the values of Eq. (\ref{al-b}), we get
$P_{\rm max}=0.7 P_{\rm lim} $.

In a similar way using Eq. \eqref{eq:Etamax}, the optimized parametrization that maximizes the efficiency of the cycle give us the value
$\eta_{\rm max}=0.34 \eta_C$.

Specific values for the maximum efficiency of the machine operating under other protocols can be obtained by substituting in Eq.~(\ref{eq:x}) the values shown in Fig.\ref{fig:optimal_ellipses_lambdaK}. This calculation shows that this machine can achieve a performance as high as $\eta_{\rm max} > 0.55 \eta_C$. 
These results are very encouraging regarding the possibility of the experimental implementation of this system.


\section{Summary and conclusions}
We have followed a geometrical approach to describe the two competing mechanisms of a non-equilibrium adiabatic thermal machine: the dissipation of energy and the heat--work conversion. While the first mechanism is described in terms of a length, the second one can be represented by and area in the parameter space. 
 We then showed that the problem of finding optimal protocols  reduces to an isoperimetric problem, which consists in finding the optimal ratio between area and length in a space with non-trivial metrics.
 
 We applied this description to a thermal machine which consists of a single qubit asymmetrically coupled  to two bosonic reservoirs at small different temperatures and driven by a cyclic protocol controlled by two parameters that vary slowly in time. 
 We solved this problem in the limit of weak coupling between the qubit and the reservoirs. 
 We analytically show the limiting value of the pumped heat between reservoirs is given by Landauer bound in an ideal Carnot cycle. We analyzed in this problem the type of cycles leading to optimal performance of the machine. Interestingly, 
 the qubit machine has a very good ratio between performance and power within a wide set of parameters. 
 
 According to our analysis, efficiencies larger 
 than $0.55$ of the Carnot cycle can be achieved and values of the corresponding output power of $0.7$ of the limiting power, corresponding to the work done in an ideal Carnot cycle divided by the duration of the cycle at which the maximum power is achieved.
 These estimates are very encouraging for 
  the experimental implementation of this machine. In this sense, a very promising platform is a superconducting qubit coupled to
  resonators, in which there are several configurations under study for some years now \cite{niskanen2003fast,mottonen2008experimental,cottet2017observing,senior2020heat,upadhyay2021robust,guthrie2021cooper}. Other possible platforms are those in which the Otto cycle has been already implemented, like
  AMO systems  \cite{abah2012nov,von2019spin,brantut2013nov}, as well as spin systems in NMR setups \cite{peterson2019dec}.
  Quantum dots, where electron pumping has been observed \cite{pothier1992jan,switkes1999adiabatic} are also candidates for implementing the heat engine and refrigerator operations as well as nanomechanical systems \cite{bachtold,ares}.
  This geometrical optimization can be also very naturally extended to analyze other systems like motors operating under slow driving and a bias voltage \cite{marun,magnet,ludovico-capone,lucas1}. In the present work we have focused on the linear-response regime, where the geometric description becomes explicit. The weak-coupling calculations of the heat and work presented in Section \ref{sec:qubit} and Appendix \ref{appA:reducedDensityMatrix} can be extended to analyze the operation beyond this regime for specific thermal machines, representing an outlook to further works.

 \section{Acknowledgements}
 
We thank Rosario Fazio and Jukka Pekola for stimulating discussions.
P.A. is supported by ``la Caixa" Foundation (ID 100010434, Grant  No.  LCF/BQ/DI19/11730023), and by the Government of Spain (FIS2020-TRANQI and Severo Ochoa CEX2019-000910-S), Fundacio Cellex, Fundacio Mir-Puig, Generalitat de Catalunya (CERCA, AGAUR SGR 1381).
 L.A. and P.T.A. are supported by CONICET, and acknowledge financial support from PIP-2015 and ANPCyT, Argentina, through PICT-2017-2726, PICT-2018-04536. L.A. thanks KITP for the hospitality in the framework of the activity "Energy and Information Transport in Non-equilibrium Quantum Systems" and the support by
 the National Science Foundation under Grant No. PHY-1748958, and  the Alexander von Humboldt Foundation, Germany. M. P.-L.  acknowledges funding from Swiss National Science Foundation through an Ambizione grant PZ00P2-186067.

\clearpage
\appendix
%
%
%
%
\section{Calculation of the linear-response coefficients}
\label{sec:linear-response}
In what follows we calculate all the linear response coefficients entering Eqs.~(\ref{w}) and~(\ref{q}) in the weak-coupling limit between the qubit and the reservoir, following Ref.~\cite{masterkeldysh}.
We rely on the adiabatic quantum-master equation approach under small temperature bias $\Delta T$ to evaluate the coefficients $\Lambda_{\mu,\nu}$, which enable the calculation of the net transferred heat and work defined in Eqs.~(\ref{w}) and~(\ref{q}). 

The derivation of the master equation corresponds to solving the non-equilibrium problem of the driven qubit coupled to the two reservoirs 
exactly up to second order in the coupling constants $V_{\alpha}^2$ and up to linear order in the velocities of the driving parameters $d_t \vec{B}$.

\subsection{Reduced density matrix}\label{appA:reducedDensityMatrix}
The Hamiltonian for the qubit can be expressed, after an appropriate unitary transformation $U$, in the instantaneous diagonal basis $|j\rangle, j=1,2$ as follows:
\begin{equation}
\label{hs1}
{\cal H}_{\rm qb}(t)=E_{1}(t) |1\rangle \langle 1| + E_{2}(t) |2\rangle \langle 2|, 
\end{equation}
where $E_{1,2}(t)=\mp B_r(t)$ are the eigenvalues of the Hamiltonian of Eq.~(\ref{qs}). We focus on the reduced density matrix $\rho(t)$ expressed in this basis, in the slow-driving regime. We  split it into a {\em frozen} plus an {\em adiabatic} contribution as 
\begin{equation}\label{split}
\rho(t) = \rho^f + \rho^a,
\end{equation}
where
the first term corresponds to the description with the Hamiltonian frozen at a given time, for which the parameters take values ${\vec B}$, while the second one corresponds to the correction $\propto d_t \vec B$.

The master equation for the corresponding matrix elements, $\rho_{ij}(t)$, reads \cite{masterkeldysh},
\begin{align}\label{eq:masterEq}
    \frac{d\rho_{ij}(t)}{dt} = \frac{i \epsilon_{ij}(t)}{h} \rho_{ij}
    +\sum_{m,n,\alpha} 
    &\left[ M^{jn}_{mi,\alpha}(t) \rho_{mn}
    + M^{in}_{jm,\alpha}(t) \rho_{nm} \right. \nonumber \\
    &\left. - M^{mn}_{jm,\alpha}(t) \rho_{in}
    - M^{mn}_{mi,\alpha}(t) \rho_{nj}
    \right]
\end{align}
where we have introduced a shorthand notation for $E_i(t)-E_j(t) =\epsilon_{ij}(t)$ for the instantaneous energy differences. The transition rates $M^{ju}_{ml,\alpha}(t)$  for the present problem are given by
\begin{equation}
\label{eq:me_coefficients}
    M^{ju}_{ml,\alpha}(t) = \frac{\xi_{\alpha, ml}(t) \xi_{\alpha, ju}(t)}{h}
    \left(
    n_\alpha(\epsilon_{ju}) \Gamma_\alpha(\epsilon_{ju}) +
    [1+n_\alpha(\epsilon_{uj})] \Gamma(\epsilon_{uj})
    \right),
\end{equation}
being $\alpha=h,c$ the reservoir indices, $n_\alpha(\varepsilon)=1/\left(e^{\varepsilon/(k_BT_{\alpha})}-1\right)$ the Bose-Einstein distribution function corresponding to the temperature of the $\alpha$-reservoir and 
$\Gamma_\alpha(\varepsilon>0) = \gamma_\alpha \varepsilon e^{-\varepsilon/\varepsilon_C}$ the bath spectral function. The functions $\xi_\alpha$ are defined for each bath as $\xi_\alpha=\hat U(t) \hat \pi_\alpha \hat U^\dagger(t)$, where $\pi_\alpha$ is defined in Eq.~\eqref{qcont}. In the present problem we use $\hat \pi_{h,c} = \hat \sigma_{x,z}$. For this problem we consider Ohmic baths with a cutoff frequency $\varepsilon_C$, and $\Gamma_\alpha(\varepsilon\leq 0)=0$. The value of $\gamma_\alpha$ depends on the coupling strength as $|V_{\alpha}|^2$. This quantity defines the relaxation time between the q-bit and the reservoirs,  $\tau_{\rm rel}\propto \gamma_{\alpha}^{-1} \varepsilon \approx \gamma_\alpha^{-1} k_B T$. 
In Eq. (\ref{eq:masterEq}) we have neglected a term proportional to
$|V_{\alpha}|^2 |d_t \vec B|$ \cite{mak3,riwar2010nov,calvo2012dec,masterkeldysh}.
This equation can be written in a compact form as
\begin{equation}
    \frac{d \mathbf{p}(t)}{dt}= \mathbf{M}(t) \mathbf{p}(t),
\end{equation}
by defining $\mathbf{p}(t)=(\rho_{11}(t),\rho_{12}(t),\rho_{21}(t),\rho_{22}(t))^T$ with the contributions $\mathbf{p}^f$ and $\mathbf{p}^a$ as in Eq. \eqref{split} and $\mathbf{M}(t)$ accordingly.

The frozen contribution $\mathbf{p}^f$ is calculated as a function of the time-dependent parameters $\vec B$ only (frozen time), and satisfies the stationary (static) limit $\frac{d \mathbf{p}^f(t)}{dt} = 0$. Hence, it can be calculated from
\begin{equation}
\label{eq:pf}
    0 = \mathbf{M}({\vec B})\cdot \mathbf{p}^f({\vec B}),
\end{equation}
with the normalization condition $\rho^f_{11}+\rho^f_{22}=1$. On the other hand, 
the adiabatic component satisfies
\begin{equation}
\label{eq:pa}
    \sum_\ell \frac{\partial \mathbf{p}^f({\vec B})}{\partial B_\ell} \dot B_\ell(t) = \mathbf{M}({\vec B})\cdot \mathbf{p^a}(t),
\end{equation}
with the normalization condition $\rho^a_{11}+\rho^a_{22}=0$. An important detail for the calculation of the partial derivatives appearing in the left side of Eq. \eqref{eq:pa} is to take into account the effects of the basis dependence in $\vec B$. A practical way to perform the derivative is by expressing $\rho^f$ in the laboratory (fixed) basis first, and then rotate back to the instantaneous diagonal basis.

We can modify $\mathbf{M}$ to include the normalization condition for $\mathbf{p}^a$ in a single equation~\cite{riwar2010nov,calvo2012dec}. Naming $\mathbf{\tilde M}$ the resulting matrix, we can finally invert Eq. (\ref{eq:pa}) to obtain the closed expression
\begin{equation}
\label{pa}
    \mathbf{p}^a(t) = \sum_n \mathbf{\tilde M}^{-1}({\vec B}) \cdot \frac{\partial \mathbf{p}^f}{\partial B_n}\dot B_n(t).
\end{equation}

\subsection{Linear-response coefficients}\label{app-lin-res}
We can now use the described approach in (\ref{appA:reducedDensityMatrix}) to obtain explicit expressions for the linear-response coefficients entering in the thermal geometric tensor $\Lambda_{\mu,\nu}$. To this end we calculate the power developed by the ac-driving sources as follows,
\begin{equation}
\label{pac}
    P_{ac}(t) = \mbox{Tr}\left[
    \dot {\cal H}_{\rm qb}(t)
    \rho(t)
    \right],
\end{equation}
being
\begin{equation}
\label{Hdot}
    \dot {\cal H}_{\rm qb}(t) = \sum_\ell 
    \frac{\partial {\cal H}_{\rm qb}(t)}{\partial B_\ell}
    \dot B_\ell(t).
\end{equation}
We now write 
\begin{equation}
    W=-\int_0^{\tau} dt  P_{ac}(t), 
\end{equation}
and replace Eq.~(\ref{split}) into Eq.~(\ref{pac}), to finally use the solutions for $\mathbf{p}^f$ and $\mathbf{p}^a$ obtained in the previous subsection. The resulting expression can be directly compared to the formal relation in Eq. \eqref{w} for $W$, where the matrix elements of the thermal geometric tensor are multiplied by different powers of $\dot B_k$.

In order to discriminate the contribution of the developed power $\propto \Delta T$,  and recalling that we are considering small temperature differences, we introduce the following extra expansion in frozen component:
${\rho}^f= {\rho}^f_T + \delta_T {\rho}^f \Delta T $. Therefore,
\begin{equation}
       \Lambda_{\ell, 3}(\vec B) =-
    \mbox{Tr} \left[
    \frac{\partial {\cal H}_{\rm qb}}{\partial B_\ell} \delta_T {\rho}^f,
    \right]. 
\end{equation}
while 
\begin{equation}
\label{eq:lambda-diss-formal}
    \Lambda_{\ell, n}(\vec B) =
    [\underline{\Lambda}]_{\ell, n}(\vec B) =
    \mbox{Tr} \left[
    \frac{\partial {\cal H}_{\rm qb}}{\partial B_\ell} 
    \left( \mathbf{\tilde M}^{-1}_T({\vec B})
    \frac{\partial \mathbf{p^f}_T}{\partial B_n} 
    \right)
    \right],
\end{equation}
where we highlight with the sub-index $T$ that the quantities are calculated with the reservoirs at the same temperature $T$ and the quantity between parentheses is to be understood as a $2x2$ matrix.
Notice that the contribution of Eq. (\ref{pac}) evaluated with ${\rho}^f_T$ corresponds to an equilibrium quantity. It represents the power developed by the conservative ac forces, and it, thus, leads to a vanishing value when averaged over the cycle.

In the same formalism used to derive the reduced density matrix, the heat current entering the reservoir $\alpha$, calculated at the second order of perturbation theory 
in the coupling to the reservoirs and within the adiabatic regime, reads \cite{calvo2012dec,masterkeldysh},
\begin{equation}
\label{Jheat}
    J_\alpha(t) = \sum_{m,n,u}
    \epsilon_{un}(t)\mbox{Re}\left[ M^{nu}_{mn,\alpha}(t) \rho_{un}(t) \right].
\end{equation}
We can follow the same logic as before to calculate this current at the first order in $\Delta T$ and $\dot{\vec B}$. The first one is 
``thermal'' component 
associated to the
the frozen components evaluated with a thermal bias $\Delta T$, while the second one is the heat current ``pumped'' by the ac driving without temperature bias.

The linear-response net transported heat between the two reservoirs is
\begin{equation}
    Q=\int_0^{\tau} dt J_{\rm c}(t)=-\int_0^{\tau} dt J_{\rm h}(t).
\end{equation}

We follow the convention of Ref. \cite{adiageo} and consistently with the definition \eqref{q}, we focus on the current entering the cold reservoir to define the net transported heat.
Associating each term of Eq. \eqref{q} with those arising from Eq. \eqref{Jheat} upon substituting ${\rho}$ by its expansion in $\Delta T$ and $\dot{\vec B}$,  we identify the linear coefficients,
\begin{equation}
\label{L3j}
    \Lambda_{3,\ell}(\vec B) =
    \vec{\Lambda}_{\ell}(\vec B) = \sum_{m,n,u}
    \epsilon_{un} 
    \mbox{Re} \left[ M^{nu}_{mn,{\rm c}, T}
    \left( \mathbf{\tilde M}_T^{-1}
    \frac{\partial \mathbf{p^f}}{\partial B_\ell} 
    \right)_{un}
    \right]
\end{equation}
and
\begin{equation}
\label{L33}
    \Lambda_{3,3}(\vec B) =
    \frac{\kappa(\vec B)}{T}
    = \sum_{m,n,u} \epsilon_{um}
    \mbox{Re} \left[  M^{nu}_{mn,{\rm c}, T}
   \delta_T \rho^f \right]_{un}
\end{equation}
These coefficients satisfy the following Onsager equations \cite{ludovico2016feb,adiageo},
\begin{equation}
\Lambda_{3,\ell}=-\Lambda_{\ell,3},\;\;\;\Lambda_{1,2}=\Lambda_{2,1}.
\end{equation}


\section{Explicit expressions for \texorpdfstring{$\underline{\Lambda}(\vec B)$}{L(B)}, \texorpdfstring{$\vec{\Lambda}(\vec B)$}{L(B)} and \texorpdfstring{$\kappa(\vec B)$}{k(B)} in the case of equal baths coupling}
\label{appendix:sameCouplingStrength}

Assuming equal spectral density in the $L$ and $R$ baths, i.e.
\begin{equation}
\begin{split}
    \Gamma_L(\epsilon) &= \bar \Gamma_L \epsilon e^{-\epsilon/\epsilon_C} =\Gamma(\epsilon) \\
    \Gamma_R(\epsilon) &= \bar \Gamma_R \epsilon e^{-\epsilon/\epsilon_C} =\Gamma(\epsilon) 
\end{split}
\end{equation}

with $\bar \Gamma_L = \bar \Gamma_R$ constants, we get explicit expressions for the complete geometric tensor. Using $(B_x, B_z) = B_r(\sin \phi, \cos \phi)$ and $\beta=1/(k_B T)$ we arrive to the following results.

\subsection*{Explicit expression for \texorpdfstring{$\underline{\Lambda}$}{L(B)}}
Using Eq. \eqref{hs1} and the expression for the reduced density matrix in the same basis, we can write for the terms with partial derivatives in Eq. \eqref{eq:lambda-diss-formal}:

\begin{equation}
\frac {\partial {\cal H}_{\rm qb}}{\partial B_\ell}=
\sum_j \partial_\ell E_j(\vec{B}) |j\rangle \langle j|
+ E_j(\vec{B}) (   |\partial_\ell j\rangle \langle j|
            +|j\rangle \langle \partial_\ell j|)
\end{equation}

\begin{equation}
    \frac{\partial \mathbf{p^f}_T}{\partial B_n} =
    \sum_{ij} \partial_\ell p^f_{T,ij}(\vec{B}) |i\rangle \langle j|
+ p^f_{T,ij}(\vec{B}) (   |\partial_\ell i\rangle \langle j|
                   +|i\rangle \langle \partial_\ell j|)
\end{equation}

with the notation $\frac{\partial}{\partial B_\ell} = \partial_\ell$. We note that the term $E_j(\vec{B})$ depends only on the absolute value of the magnetic field $B_r$. In addition, the density matrix $\mathbf{p}^f_T(\vec{B})$ is a function of the energy spectrum, since it is computed in thermal equilibrium considering both baths at equal temperature $T$, and thus depends only on $B_r$ as well.

On the other hand, the operators $|\partial_\ell i\rangle \langle j|$ and $|i\rangle \langle \partial_\ell j|$ are nonzero only for variations in the polar coordinate $\phi$ because the eigenvectors $|i\rangle$ are associated to the unitary transformation that makes ${\cal H}_{\rm qb}$ diagonal, and do not change when $\vec{B}$ stays in the same direction.

These facts allows us to separate the linear response coefficient $\underline{\Lambda}$ into a radial contribution ${\lambda}_{r}$ (changes in the energy spectrum) and a polar contribution ${\lambda}_{\phi}$ (basis rotation). Inserting the solution to \eqref{eq:pf} and \eqref{pa} into Eq. \eqref{eq:lambda-diss-formal} we get:

\begin{equation}
\label{eq:lambda_diss_radial}
\langle r| \underline{\Lambda} |r \rangle =
{\lambda}_{r}(\vec B)=
\frac{\hbar \beta \text{sinh}(\beta  B_r)}{\Gamma (2 B_r) \text{cosh}^3(\beta  B_r)}
\end{equation}

\begin{equation}
\label{eq:lambda_diss_polar}
\langle \phi| \underline{\Lambda} |\phi \rangle =
{\lambda}_{\phi}(\vec B)=
\frac{\hbar \Gamma (2 B_r)}{4 B_r^3}
\end{equation}

Eq. \eqref{eq:lambda-diss-formal} finally reads:

\begin{equation}
\underline{\Lambda} =
\lambda_{r}|r\rangle \langle r|+
\lambda_{\phi}|\phi\rangle \langle \phi|.
\end{equation}

\begin{figure}
    \centering
    \includegraphics[width=0.45\textwidth]{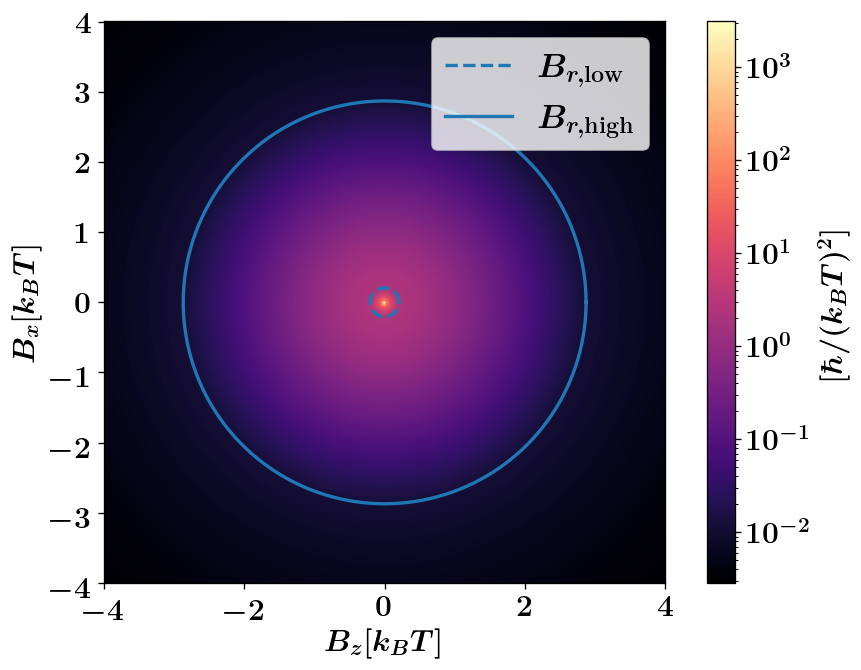}
    \includegraphics[width=0.45\textwidth]{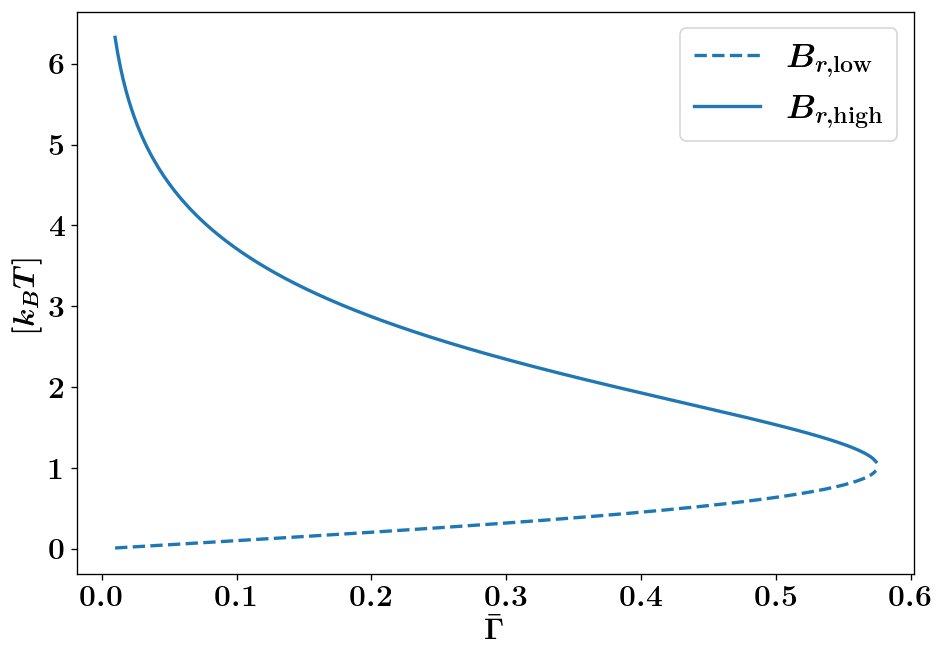}
    \caption{Upper panel: Maximum eigenvalue $\mbox{max}\left[\lambda_{r}, \lambda_{\phi}\right]$, which ultimately defines the maximum possible dissipation at a given point $\vec B$.
    Right panel: $B_{r,\rm low}$ (dash line) and $B_{r,\rm high}$ (solid line) as function of $\bar\Gamma$. The values $B_r$ for which $\lambda_{r}=\lambda_{\phi}$ are determined by $\bar \Gamma$ only and scales linearly with $k_B T$.}
    \label{fig:dissipation_change}
\end{figure}

We now turn to analyze $L^2$, which quantifies the dissipated energy for a particular protocol, determined by $\underline{\Lambda}$ through Eq. \eqref{eq:geometric_k}. In the upper panel of Fig. \ref{fig:dissipation_change} we show the maximum eigenvalue, $\mbox{max}\left[\lambda_{r}, \lambda_{\phi}\right]$ as a function of $(B_z,B_x)$.
This plot reflects the behavior resulting from the analytical expressions of Eqs. \eqref{eq:lambda_diss_radial} and \eqref{eq:lambda_diss_polar}. 
At every point, and depending on the values of $\bar \Gamma$ and $T$, this maximum eigenvalue corresponds to a pure polar or pure radial displacement of $\vec{B}$.
Within the small circle plotted in dashed lines and outside the one in solid lines, the highest eigenvalue is $\lambda_{\phi}$, while within the two circles, $\lambda_{r}$ is the largest one.
This means that for small $B_r < B_{r,\rm low}$ as well as for large $B_r > B_{r, \rm high}$, protocols leading to the smallest dissipation are those associated to changes in $B_r$, while 
for $B_{r,\rm low}< B_r < B_{r, \rm high}$, protocols associated to rotations are the least dissipative ones.
The specific values  $B_{r,\rm low},\; B_{r,\rm high}$
depend on the temperature and the coupling between the qubit and the reservoirs, as shown in Fig.~\ref{fig:dissipation} for two values of $\Gamma$.

The precise shape of the interval $B_{r,\rm low},\; B_{r,\rm high}$ is shown in the lower panel of Fig.~\ref{fig:dissipation_change} and depends only on $\bar \Gamma$, while the final value has a linear dependence on $k_B T$. We see that for $\bar\Gamma \gtrapprox 0.6$ the rotational dissipation dominates for all values of $|\vec B|$.

\subsection*{Explicit expression for \texorpdfstring{$\vec \Lambda$}{L} and \texorpdfstring{$\kappa$}{k}}
Again, plugging the expressions describing $\mathbf{p}(t)$ given by Eqs. \eqref{eq:pf} and \eqref{pa} into Eq. \eqref{L3j} we get for the qbit:

\begin{equation}
    \vec \Lambda_{1}(\vec B) = -\beta  B_r \sin ^3(\phi ) \text{sech}^2(\beta  B_r)
\end{equation}
\begin{equation}
    \vec \Lambda_{2}(\vec B) = -\beta  B_r \sin ^2(\phi ) \cos (\phi ) \text{sech}^2(\beta  B_r).
\end{equation}

And lastly, using Eq. \eqref{L33}, the thermal conductance is explicitly written as:

\begin{equation}
    \kappa(\vec B) = \frac{4 B_r^2 \sin ^2( \phi ) \cos ^2( \phi ) \Gamma (2 B_r) \text{csch}\left(\frac{2 B_r}{T k_B}\right)}{\hbar k_B T}.
\end{equation}
\section{More discussion on the circular sector}
\label{app:pizzaProtocol}
\begin{figure}[h]
    \centering
    \includegraphics[width=0.5\textwidth]{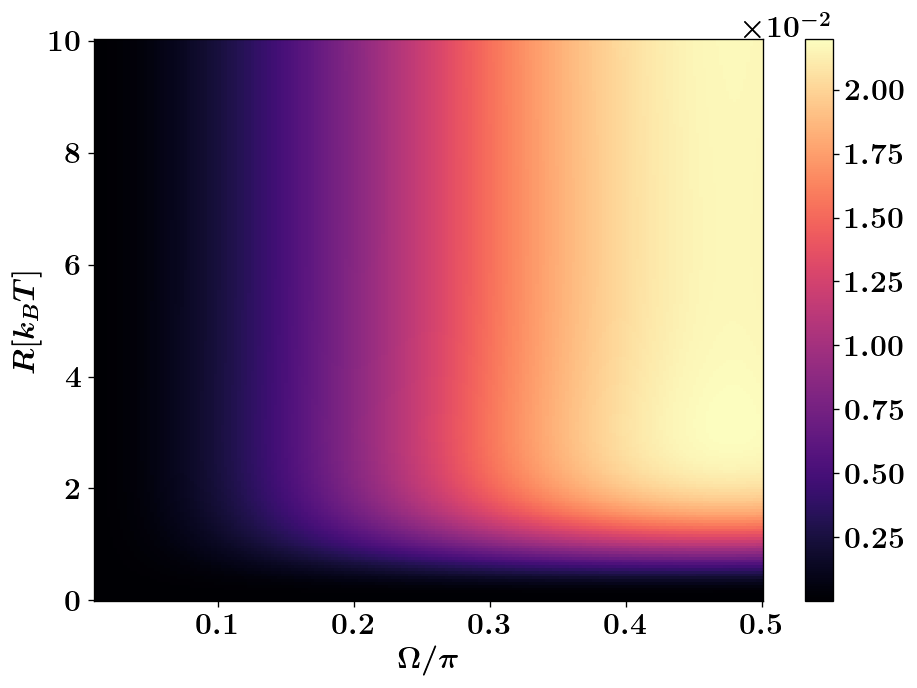}
    \caption{Geometrical values $A^2/\mathcal{L}^2$ for the circular-sector protocols. We found a saturation value equal to $0.022 (k_B T)^2/\hbar$ which corresponds to the trajectory defined by $\Omega=\pi/2$ and $R \rightarrow \infty$.}
    \label{fig:pizza_power}
\end{figure}

Here we present a more detailed study on the power and the efficiency of the circular sector defined in section \ref{sec:results-C}. In Fig. \ref{fig:pizza_power} we compute the geometrical value $A^2/\mathcal{L}^2$ as a function of the two parameters $R$ and $\Omega$ that define the curves of this class.

We see from Fig. \ref{fig:pizza_power} that for big enough $R$ the value of $A^2/\mathcal{L}^2$ depends only on $\Omega$. This fact can be understood by looking at the upper panel of Fig. \ref{fig:dissipation_change}: for $B_r>B_{r,{\rm high}}$ we have $\underline{\Lambda} \approx 0$, which implies that the dissipation outside the solid circle is negligible, and only the radial sections contained in the solid circle contributes significantly to $L^2$. Furthermore, since $\underline{\Lambda}$ has rotational symmetry, the value of $L^2$ is independent of the direction of the radial parts. These two observations leads us to the conclusion that $\mathcal{L}^2$ has a saturation value in the circular sectors with $R>>B_{r,{\rm high}}$. Finally, the dependence on $\Omega$ is explained by looking at Fig. \ref{fig:curl_lambda}, where it is clear that $A=A(\Omega)$ for $R>>B_{r,{\rm high}}$ as well, and the maximum occurs at $\Omega=\pi/2$. The saturation value of $A^2/\mathcal{L}^2$ for the circular sector is found to be $\max_{\rm circ-sec} \frac{A^2}{\mathcal{L}^2} = 0.022 (k_B T)^2/\hbar$, which comparing to Fig. \ref{fig:optimal_ellipses} is around $30\%$ smaller than the ellipses case.

\begin{figure}[ht]
    \centering
    \includegraphics[width=0.5\textwidth]{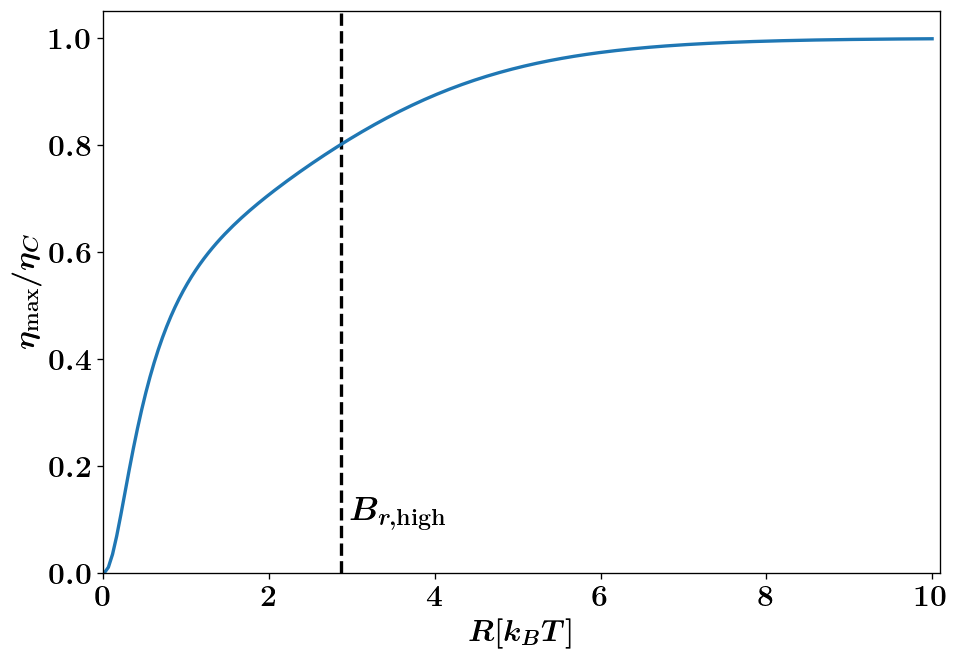}
    \caption{Computed efficiency for the circular sector with $\Omega = \pi/2$ as a function of the $R$ parameter. At $R \rightarrow \infty$ we recover Carnot efficiency for the optimal parametrization.}
    \label{fig:pizza_efficiency}
\end{figure}

In Fig. \ref{fig:pizza_efficiency} we show the computed maximum efficiency $\eta_{\rm max}$ in Eq. \eqref{eq:Etamax} of the circular-sector protocol with $\Omega=\pi/2$ as a function of $R$.

Recall from discussion in section~\ref{sec:max_efficiency} that the limiting case of the circular sectors with the mentioned $\Omega$ and $R\rightarrow\infty$ defines a Carnot cycle when optimized for maximum efficiency. This fact is clearly seen in Fig.~\ref{fig:pizza_efficiency} as the saturation value of $\eta_{max}/\eta_C$ goes to $1$ as $R >> B_{r,{\rm high}}$.

\bibliography{opt-geo.bib}
\end{document}